\documentclass[preprint,authoryear]{elsarticle}

\usepackage{cite}
\usepackage{verbatim}
\usepackage{color}
\usepackage{colortbl}
\usepackage{setspace}
\usepackage{pdflscape}
\usepackage{appendix}
\usepackage{graphicx}
\usepackage{epstopdf}
\usepackage{hyperref}
\usepackage[center]{subfigure}

\definecolor{red}{rgb}{0.5,0,0}
\definecolor{blue}{rgb}{0,0,0.5}
\definecolor{black}{rgb}{0,0,0}

\begin{document}

\begin{frontmatter}

\title{Breaking Monotony with Meaning: \\ Motivation in Crowdsourcing Markets\tnoteref{acks}}

\author[bcuc]{Dana Chandler\corref{cor1}}
\ead{dchandler@mit.edu}
\author[wsupenn]{Adam Kapelner}
\ead{kapelner@wharton.upenn.edu}

\cortext[cor1]{Principal corresponding author}

\address[bcuc]{Massachusetts Institute of Technology \\ 50 Memorial Drive \\ Cambridge, MA 02142} \vspace{1pt}
\address[wsupenn]{The Wharton School of the University of Pennsylvania, \\3730 Walnut Street \\Philadelphia, PA 19104}

\tnotetext[acks]{Both authors contributed equally to this work. The authors wish to thank Professor Susan Holmes of Stanford University for comments and for allowing us to adapt the DistributeEyes software for our experiment (funded under NIH grant \#R01GM086884-02). They gratefully acknowledge financial support from the George and Obie Schultz Fund and the NSF Graduate Research Fellowship Program. The authors also thank Iwan Barankay, Lawrence Brown, Rob Cohen, Geoff Goodwin, Patrick DeJarnette, John Horton, David Jimen\'ez-Gomez, Emir Kamenica, Abba Krieger, Steven Levitt, Blakeley McShane, Susanne Neckermann, Paul Rozin, Martin Seligman, Jesse Shapiro, J\"{o}rg Spenkuch, Jan Stoop, Chad Syverson, Mike Thomas, Abraham Wyner, seminar participants at the University of Chicago, and our reviewers.}

\begin{abstract}
We conduct the first natural field experiment to explore the relationship between the ``meaningfulness'' of a task and worker effort. We employed about 2,500 workers from Amazon's Mechanical Turk (MTurk), an online labor market, to label medical images. Although given an identical task, we experimentally manipulated how the task was framed. Subjects in the \textit{meaningful} treatment were told that they were labeling tumor cells in order to assist medical researchers, subjects in the \textit{zero-context} condition (the control group) were not told the purpose of the task, and, in stark contrast, subjects in the \textit{shredded} treatment were not given context and were additionally told that their work would be discarded. We found that when a task was framed more meaningfully, workers were more likely to participate. We also found that the meaningful treatment increased the quantity of output (with an insignificant change in quality) while the shredded treatment decreased the quality of output (with no change in quantity). We believe these results will generalize to other short-term labor markets. Our study also discusses MTurk as an exciting platform for running natural field experiments in economics.
\end{abstract}


\end{frontmatter}

\section{Introduction}\label{sec-intro}

Economists, philosophers, and social scientists have long recognized that non-pecuniary factors are powerful motivators that influence choice of occupation. For a multidisciplinary literature review on the role of meaning in the workplace, we recommend \citet{Rosso2010}. Previous studies in this area have generally been based on ethnographies, observational studies, or laboratory experiments. For instance, \citet{Wrzesniewski1997} used ethnographies to classify work into jobs, careers, or callings. Using an observation study, \citet{preston_1989} demonstrated that workers may accept lower wages in the non-profit sector in order to produce goods with social externalities. Finally, \citet{Ariely_search_for_meaning_2008} showed that labor had to be both recognizable and purposeful to have meaning. In this paper, we limit our discussion to the role of meaning in economics, particularly through the lens of competing differentials. We perform the first \textit{natural field experiment} \citep{Harrison_List_natural_field_2004} in a real effort task that manipulates levels of meaningfulness. This method overcomes a number of shortcomings of the previous literature, including: interview bias, omitted variable bias, and concerns of external validity beyond the laboratory.

We study whether employers can deliberately alter the perceived ``meaningfulness'' of a task in order to induce people to do more and higher quality work and thereby work for a lower wage. We chose a task that would appear meaningful for many people if given the right context --- helping cancer researchers mark tumor cells in medical images. Subjects in the \textit{meaningful} treatment were told the purpose of their task is to ``help researchers identify tumor cells;'' subjects in our \textit{zero-context} group were not given any reason for their work and the cells were instead referred to as mere ``objects of interest'' and laborers in the \textit{shredded} group were given zero context but also explicitly told that their labelings would be discarded upon submission. Hence, the pay structure, task requirements, and working conditions were identical, but we added cues to alter the perceived meaningfulness of the task.

We recruited workers from the United States and India from Amazon's Mechanical Turk (MTurk), an online labor market where people around the world complete short, ``one-off'' tasks for pay.  The MTurk environment is a spot market for labor characterized by relative anonymity and a lack of strong reputational mechanisms. As a result, it is well-suited for an experiment involving the meaningfulness of a task since the variation we introduce regarding a task's meaningfulness is less affected by desires to exhibit pro-social behavior or an anticipation of future work (career concerns). We ensured that our task appeared like any other task in the marketplace and was comparable in terms of difficulty, duration, and wage.

Our study is representative of the kinds of natural field experiments for which MTurk is particularly suited. Section \ref{subsec:mturk_for_nfes} explores MTurk's potential as a platform for field experimentation using the framework proposed in \citet{Levitt_laboratory_experiments_2007,levitt2009field}.

We contribute to the literature on compensating wage differentials \citep{rosen1986theory} and the organizational behavioral literature on the role of meaning in the workplace \citep{Rosso2010}. Within economics, \citet{stern2004scientists} provides quasi-experimental evidence on compensating differentials within the labor market for scientists by comparing wages for academic and private sector job offers among recent Ph.D. graduates. He finds that ``scientists pay to be scientists'' and require higher wages in order to accept private sector research jobs because of the reduced intellectual freedom and a reduced ability to interact with the scientific community and receive social recognition. \citet{Ariely_search_for_meaning_2008} use a laboratory experiment with undergraduates to vary the meaningfulness of two separate tasks: (1) assembling Legos and (2) finding 10 instances of consecutive letters from a sheet of random letters. Our experiment augments experiment 1 in \citet{Ariely_search_for_meaning_2008} by testing whether their results extend to the field. Additionally, we introduce a richer measure of task effort, namely \textit{task quality}. Where our experiments are comparable, we find that our results parallel theirs.

We find that the main effects of making our task more meaningful is to induce a higher fraction of workers to complete our task, hereafter dubbed as ``induced to work.'' In the meaningful treatment, 80.6\% of people labeled at least one image compared with 76.2\% in the zero-context and 72.3\% in the shredded treatments. 

After labeling their first image, workers were given the opportunity to label additional images at a declining piecerate. We also measure whether the treatments increase the quantity of images labeled. We classify participants as ``high-output'' workers if they label five or more images (an amount corresponding to roughly the top tercile of those who label) and we find that workers are approximately 23\% more likely to be high-output workers in the meaningful group.

We introduce a measure of task quality by telling workers the importance of accurately labeling each cell by clicking as close to the center as possible. We first note that MTurk labor is high quality, with an average of 91\% of cells found. The meaning treatment had an ambiguous effect, but the shredded condition in both countries lowered the proportion of cells found by about 7\%.

By measuring both quantity and quality we are able to observe how task effort is apportioned between these two ``dimensions of effort.'' Do workers work ``harder'' or ``longer'' or both? We found an interesting result: the meaningful condition seems to increase quantity without a corresponding increase in quality and the shredded treatment decreases quality without a corresponding decrease in quantity. Investigating whether this pattern generalizes to other domains may be a fruitful future research avenue.

Finally, we calculate participants' average hourly waged based on how long they spent on the task. We find that subjects in the meaningful group work for \$1.34 per hour, which is 6 cents less per hour than zero context participants and 14 cents less per hour than shredded condition participants.

We expect our findings to generalize to other short-term work environments such as temporary employment or piecework. In these environments, employers may not consider that non-pecuniary incentives of meaningfulness matter; we argue that these incentives do matter, and to a significant degree.

Section \ref{sec-background} provides background on MTurk and discusses its use as a platform for conducting economic field experiments. Section \ref{sec-design} describes our experimental design. Section \ref{sec:results} presents our results and discussion and Section \ref{sec-conclusion} concludes. \ref{sec:exp_design_appendix} provides full details on our experimental design and \ref{sec:technical_discussion} is a technical appendix for conducting experiments using the MTurk platform.

\section{Mechanical Turk and its potential for field experimentation}\label{sec-background}

Amazon's Mechanical Turk (MTurk) is the largest online, task-based labor market and is used by hundreds of thousands of people worldwide. Individuals and companies can post tasks (known as Human Intelligence Tasks, or ``HITs'') and have them completed by an on-demand labor force. Typical tasks include image labeling, audio transcription, and basic internet research. Academics also use MTurk to outsource low-skilled resource tasks such as identifying linguistic patterns in text \citep{sprouse2011validation} and labeling medical images \citep{Holmes_distributeyes_2010}. The image labeling system from the latter study, known as ``DistributeEyes,''  was originally used by breast cancer researchers and was modified for our experiment.

Beyond simply using MTurk as a source of labor, academics have also began using MTurk as a way to conduct online experiments. The remainder of the section highlights some of the ways this subject pool is used and places special emphasis on the suitability of the environment for natural field experiments in economics.

\subsection{General use by social scientists}

As \citet{henrich2010weirdest} argue, many findings from social science are disproportionately based on what he calls ``W.E.I.R.D.'' subject pools (\textbf{W}estern, \textbf{E}ducated, \textbf{I}ndustrialized, \textbf{R}ich, and \textbf{D}emocratic) and as a result it is inappropriate to believe the results generalize to larger populations. Since MTurk has users from around the world, it is also possible to conduct research across cultures. For example, \citet{eriksson2010emotional} use a cross-national sample from MTurk to test whether differential preferences for competitive environments are explained by females' stronger emotional reaction to losing, hypothesized by \citet{croson2009gender}.

It is natural to ask whether results from MTurk generalize to other populations. \citet{paolacci2010running} assuage these concerns by replicating three classic framing experiments on MTurk: The Asian Disease Problem, the Linda Problem and the Physician Problem; \citet{horton_onlinelab_2010} provide additional replication evidence for experiments related to framing, social preferences, and priming. \citet{Berinsky2012} argues that the MTurk population has ``attractive characteristics'' because it approximates gold-standard probability samples of the US population. All three studies find that the direction and magnitude of the effects line up well compared with those found in the laboratory. 

An advantage of MTurk relative to the laboratory is that the researcher can rapidly scale experiments and recruit hundreds of subjects within only a few days and at substantially lower costs.\footnote{For example, in our study we paid 2,471 subjects \$789 total and they worked 701 hours (equating to 31 cents per observation). This includes 60 subjects whose data were not usable.}

\subsection{Suitability for natural field experiments in Economics}\label{subsec:mturk_for_nfes}

Apart from general usage by academics, the MTurk environment offers additional benefits for experimental economists and researchers conducting natural field experiments. We analyze the MTurk environment within the framework laid out in \citet{Levitt_laboratory_experiments_2007,levitt2009field}. 

In the ideal natural field experiment, ``the environment is such that the subjects naturally undertake these tasks and [do not know] that they are participants in an experiment.'' Additionally, the experimenter must exert a high degree of control over the environment without attracting attention or causing participants to behave unnaturally. MTurk's power comes from the ability to construct customized and highly-tailored environments related to the question being studied. It is possible to collect very detailed measures of user behavior such as precise time spent on a webpage, mouse movements, and positions of clicks. In our experiment, we use such data to construct a precise quality measure.

MTurk is particularly well-suited to using experimenter-as-employer designs \citep{list_experimenter_as_employer_2006} as a way to study worker incentives and the employment relationship without having to rely on cooperation of private sector firms.\footnote{\citet{barankay2010rankings}  remarks that ``the experimenter [posing] as the firm [gives] substantial control about the protocol and thereby eliminates many project risks related to field experiments.} For example, \citet{barankay2010rankings} posted identical image labeling tasks and varied whether workers were given feedback on their relative performance (i.e., ranking) in order to study whether providing rank-order feedback led workers to return for a subsequent work opportunity.  For a more detailed overview of how online labor markets can be used in experiments, see \citet{horton_onlinelab_2010}.

\citet{Levitt_laboratory_experiments_2007} enumerate possible complications that arise when experimental findings are extrapolated outside the lab: \textit{scrutiny}, \textit{anonymity}, \textit{stakes}, \textit{selection}, and \textit{artificial restrictions}.  We analyze each complication in the context of our experiment and in the context of experimentation using MTurk in general.

\textit{Scrutiny} and \textit{anonymity}. In the lab, experimenter effects can be powerful; subjects behave differently if they are aware their behavior is being watched. Relatedly, subjects frequently lack anonymity and believe their choices will be scrutinized after the experiment. In MTurk, interaction between workers and employers is almost non-existent; most tasks are completed without any communication and workers are only identifiable by a numeric identifier. Consequently, we believe that MTurk experiments are less likely to be biased by these complications.

\textit{Stakes}. In the lab or field, it's essential to ``account properly for the differences in stakes across settings'' \citep{Levitt_laboratory_experiments_2007}. We believe that our results would generalize to other short-term work environments, but would not expect them to be generalizable to long-term employment decisions such as occupational choice. Stakes must also be chosen adequately for the environment and so we were careful to match wages to the market average.

\textit{Selection}. Experiments fail to be generalizable when ``participants in the study differ in systematic ways from the actors engaged in the targeted real-world setting.'' We know that within MTurk, it is unlikely that there is selection into our experiment since our task was designed similar in appearance to real tasks. The MTurk population also seems representative along a number of observable demographic characteristics \citep{Berinsky2012}; however, we acknowledge that there are potentially unobservable differences between our subject pool and the broader population. Still, we believe that MTurk subject behavior would generalize to workers' behavior in other short-term labor markets. 

\textit{Artificial restrictions}. Lab experiments place unusual and artificial restrictions on the actions available to subjects and they examine only small, non-representative windows of time because the experimenter typically doesn't have subjects and time horizons for an experiment. In structuring our experiment, workers had substantial latitude in how they performed their task. In contrast with the lab, subjects could ``show-up'' to our task whenever they wanted, leave at will, and were not time-constrained. Nevertheless, we acknowledge that while our experiment succeeded in matching short-term labor environments like MTurk, that our results do not easily generalize to longer-term employment relationships.

\citet{levitt2009field} highlight two limitations of field experiments vis-a-vis laboratory experiments: the \textit{need for cooperation} with third parties and the difficulty of \textit{replication}. MTurk does not suffer from these limitations. Work environments can be created by researchers without the need of a private sector partner, whose interests may diverge substantially from that of the researcher. Further, MTurk experiments can be replicated simply by downloading source code and re-running the experiment. In many ways, this allows a “push-button’ replication that is far better than that offered in the lab.

\section{Experimental Design}\label{sec-design}

\subsection{Subject recruitment}

In running our randomized natural field experiment, we posted our experimental task so that it would appear like any other task (image labeling tasks are among the most commonly performed tasks on MTurk). Subjects had no indication they were participating in an experiment. Moreover, since MTurk is a market where people ordinarily perform one-off tasks, our experiment could be listed inconspicuously. 

We hired a total of 2,471 workers (1,318 from the US and 1,153 from India). Although we tried to recruit equally from both countries, there were fewer Indians in our sample since attrition in India was higher. We collected each worker's age and gender during a ``colorblindness'' test that we administered as part of the task. These and other summary statistics can be found in Table \ref{tab:summary_stats}. By contracting workers from the US and India, we can also test whether workers from each country respond differentially to the meaningfulness of a task. 

Our task was presented so that it appeared like a one-time work opportunity (subjects were barred from doing the experiment more than once) and our design sought to maximize the amount of work we could extract during this short interaction. The first image labeling paid \$0.10, the next paid \$0.09, etc, leveling off at \$0.02 per image. This wage structure was also used in \citet{Ariely_search_for_meaning_2008} and has the benefit of preventing people from working too long.

\subsection{Description of experimental conditions}

Upon accepting our task, workers provided basic demographic information and passed a color-blindness test. Next, they were randomized into either the \textit{meaningful}, the \textit{zero-context}, or the \textit{shredded} condition. Those in the shredded condition were shown a warning message stating that their labeling will not be recorded and we gave them the option to leave. Then, all participants were forced to watch an instructional video which they could not fast-forward. See \ref{sec:exp_design_appendix} for the full script of the video as well as screenshots.

The video for the meaningful treatment began immediately with cues of meaning. We adopt a similar working definition of ``meaningfulness'' as used in \citet{Ariely_search_for_meaning_2008}: ``Labor [or a task] is meaningful to the extent that (a) it is recognized and/or (b) has some point or purpose.''

We varied the levels of meaningfulness by altering the degree of recognition and the detail used to explain the purpose of our task. In our meaningful group, we provided ``recognition'' by thanking the laborers for working on our task. We then explained the ``purpose'' of the task by creating a narrative explaining how researchers were inundated with more medical images than they could possibly label and that they needed the help of ordinary people. In contrast, the zero-context and shredded groups were not given recognition, told the purpose of the task, or thanked for participating; they were only given basic instructions. Analyzing the results from a post-manipulation check (see section \ref{subsec:pmc}), we are confident that these cues of meaning induced the desired affect.

Both videos identically described the wage structure and the mechanics of how to label cells and properly use the task interface (including zooming in/out and deleting points, which are metrics we analyze). However, in the meaningful treatment, cells were referred to as ``cancerous tumor cells'' whereas in the zero-context and shredded treatments, they were referred to as nondescript ``objects of interest.'' Except for this phrase change, both scripts were identical during the instructional sections of the videos. To emphasize these cues, workers in the meaningful group heard the words ``tumor,'' ``tumor cells,'' ``cells,'' etc. 16 times before labeling their first image and similar cues on the task interface reminded them of the purpose of the task as they labeled.

\subsection{Task interface, incentive structure, and response variables}
        
After the video, we administered a short multiple-choice quiz testing workers' comprehension of the task and user interface. In the shredded condition, we gave a final question asking workers to again acknowledge that their work will not be recorded. 

\begin{figure}[htp]
\centering
\includegraphics[width=3.856in]{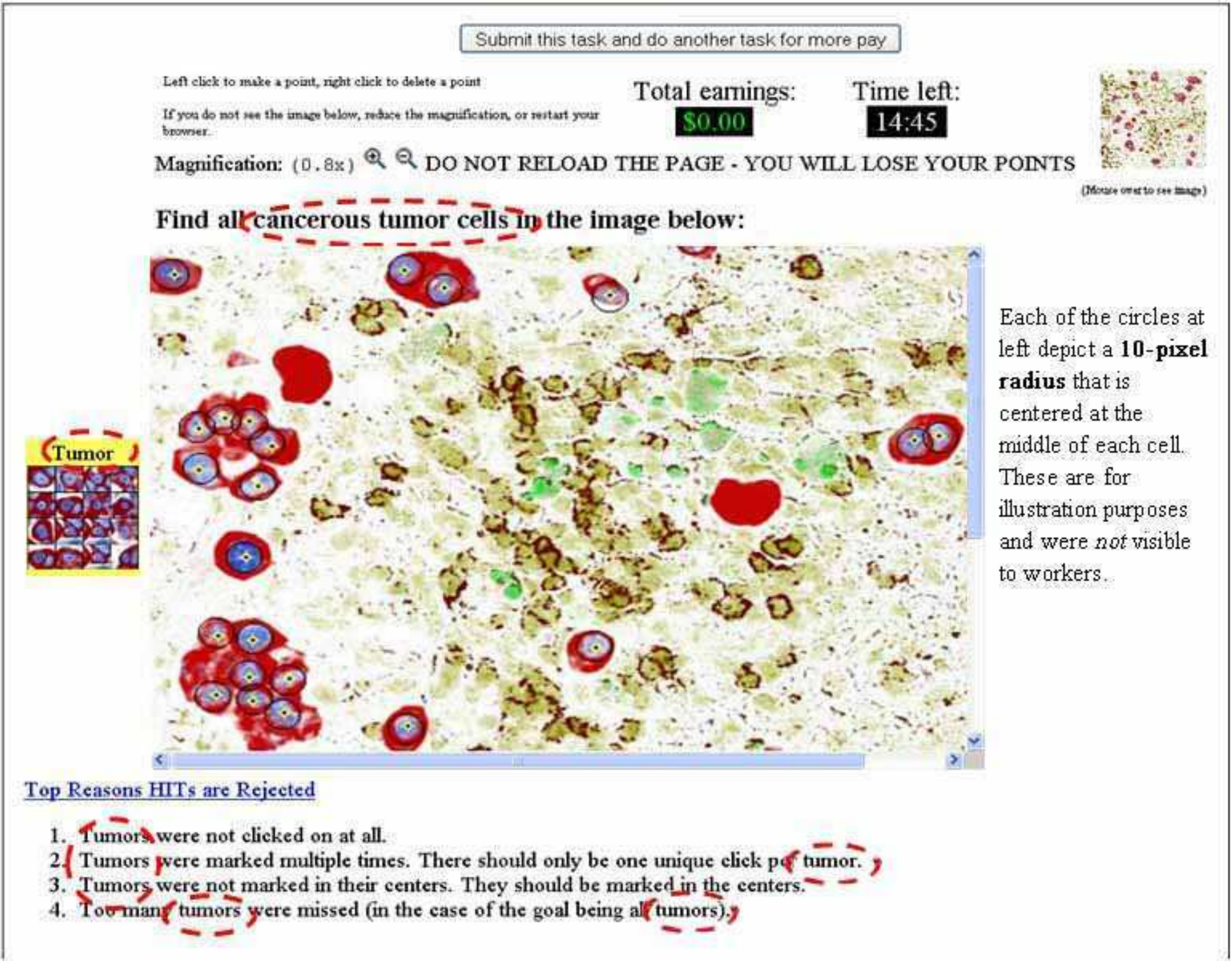}
\caption{{\bf Main task portal for a subject in the meaningful treatment} Workers were asked to identify all tumors in the image. Each image had 90 cells and took 5 minutes on average. Our interface reminds the workers in 8 places that they are identifying tumor cells. The black circles around each point were \textit{not} visible to participants. We display them to illustrate the size of a 10-pixel radius.}
\label{fig:interface}
\end{figure}

Upon passing the quiz, workers were directed to a task interface which displayed the image to be labeled and allowed users to mark cancerous tumor cells (or ``objects of interest'') by clicking (see figure \ref{fig:interface}). The image shown was one of ten look-alike photoshopped images displayed randomly. We also provide the workers with controls --- \textit{zoom functionality} and the ability to \textit{delete points} --- whose proper use would allow them to produce high-quality labelings. 

During the experiment, we measured three response variables: (1) induced to work, (2) quantity of image labelings, and (3) quality of image labelings. 

Many subjects can -- and -- do stop performing a task even after agreeing to complete it. While submitting bad work on MTurk is penalized, workers can abandon a task with only nominal penalty. Hence, we measure attrition with the response variable \textit{induced to work}. Workers were only counted as induced to work if they watched the video, passed the quiz, and completed one image labeling. Our experimental design deliberately encourages attrition by imposing an upfront and unpaid cost of watching a three-minute instructional video and passing a quiz before moving on to the actual task. 

Workers were paid \$0.10 for the first image labeling. They were then given an option to label another image for \$0.09, and then another image for \$0.08, and so on.\footnote{Each image was randomly picked from a pool of ten look-alike images.} At \$0.02, we stopped decreasing the wage and the worker was allowed to label images at this pay rate indefinitely. After each image, the worker could either collect what they had earned thus far, or label more images. We used the \textit{quantity of image labelings} for our second response variable.

In our instructional video, we emphasized the importance of marking the exact center of each cell. When a worker labeled a cell by clicking on the image, we measured that click location to the nearest pixel. Thus, we were able to detect if the click came ``close'' to the actual cell. Our third response variable, \textit{quality of image labelings} is the proportion of objects identified based on whether a worker's click fell within a pixel radius from the object's true center. We will discuss the radii we picked in the following section.

After workers chose to stop labeling images and collect their earnings, they were given a five-question PMC survey which asked whether they thought the task (a) was enjoyable (b) had purpose (c) gave them a sense of accomplishment (d) was meaningful (e) made their efforts recognized. Responses were collected on a five-point Likert scale. We also provided a text box to elicit free-response comments.\footnote{About 24\% of respondents left comments (no difference across treatments).}

\subsection{Hypotheses}

\paragraph{Hypothesis 1} We hypothesize that at equal wages, the meaningful treatment will have the highest proportion of workers induced to work and the shredded condition will have the lowest proportion. In the following section, we provide theoretical justification for this prediction.

\paragraph{Hypothesis 2} As in \citet{Ariely_search_for_meaning_2008}, we hypothesize that \textit{quantity} of images labeled will be increasing in the level of meaningfulness.

\paragraph{Hypothesis 3} In addition to quantity, we measure the \textit{quality} of image labelings and hypothesize that this is increasing in the level of meaningfulness.

\paragraph{Hypothesis 4} Based upon prior survey research on MTurk populations, we hypothesize that \textit{Indian workers are less responsive to meaning}. \citet{panos_demographics_2010} finds that Indians are more likely to have MTurk as a primary source of income (27\% vs. 14\% in the US). Likewise, people in the US are nearly twice as likely to report doing tasks because they are fun (41\% vs. 20\%). Therefore, one might expect financial motivations to be more important for Indian workers.\footnote{Although \citet{horton_onlinelab_2010} find that workers of both types are strongly motivated by money.}

\section{Experimental Results and Discussion}\label{sec:results}

We ran the experiment on $N=2,471$ subjects (1,318 from the United States and 1,153 from India). Table \ref{tab:summary_stats} shows summary statistics for our response variables (induced to work, number of images, and quality), demographic variables, and hourly wage.

\begin{table}[htbp]
\centering\medskip
\small
\begin{tabular}{rlllll}
  & Shredded & Zero  & Meaningful     & US  & India  \\ 
  &  & Context  &      & only  & only  \\ \hline  \hline 
\% Induced to Work & .723 & .762 & .806 & .85 & .666 \\
\# Images (if $\geq$ 1) & 5.94 $\pm$ 6.8 & 6.11 $\pm$ 6.9  & 7.12 $\pm$ 7.6 & 5.86 $\pm$ 6.1 & 7.17 $\pm$ 8.3 \\ 
Did $\geq$ 2 labelings & .696 & .706 & .75 & .797 & .627 \\ 
Did $\geq$ 5 labelings & .343 & .371 & .456 & .406 & .373 \\  
Avg Hourly Wage & \$1.49 & \$1.41 & \$1.34 & \$1.50 & \$1.29 \\ 
\% Male & .616 & .615 & .58 & .483 & .743 \\ 
Age & 29.6 $\pm$ 9.3 & 29.6 $\pm$ 9.5 & 29.3 $\pm$ 9.1 & 31.8 $\pm$ 10.5 & 26.9 $\pm$ 6.8  \\ 
$N$ & 828 & 798 & 845 & 1318 & 1153 \\ \hline 
Coarse quality & .883 $\pm$ .21 & .904 $\pm$ .18 & .930 $\pm$ .14 & .924 $\pm$ .15 & .881 $\pm$ .21 \\ 
Fine quality & .614 $\pm$ .22  & .651 $\pm$ .21 & .676 $\pm$ .18 & .668 $\pm$ .19 & .621 $\pm$ .26 \\ 
PMC Meaning & 3.44 $\pm$ 1.3  & 3.54 $\pm$ 1.2 & 4.37 $\pm$ 0.9 & 3.67 $\pm$ 1.3 & 3.98 $\pm$ 1.1 \\  \hline
\end{tabular}
\caption{Summary statistics for response variables and demographics by treatment and country. The statistics for the quality metrics are computed by averaging each worker's average quality (only for workers who labeled one or more images). The statistics for the PMC meaning question only include workers who finished the task and survey.}
\label{tab:summary_stats}
\end{table}

Broadly speaking, as the level of meaning increases, subjects are more likely to participate and they label more images and with higher quality. Across all treatments, US workers participate more often, label more images, and mark points with greater accuracy. Table \ref{tab:effects_summary} uses a heatmap to illustrate our main effect sizes and their significance levels by treatment, country, and response variable. Each cell indicates the size of a treatment effect relative to the control (i.e., zero context condition). Statistically significant \textit{positive} effects are indicated using green fill where darker green indicates higher levels of significance. Statistically significant \textit{negative} effects are indicated using red fill where darker red indicates higher levels of significance. Black text without fill indicates effects that are marginally significant ($p < 0.10$). Light gray text indicates significance levels above 0.10.

Overall, we observe that the meaningful condition induces an increase in quantity without significantly increasing quality, and the shredded condition induces a quality decrease with quantity remaining constant. This ``checkerboard effect'' may indicate that meaning plays a role in moderating how workers trade quantity for quality i.e. how their energy is channeled in the task.

We now investigate each response variable individually.

\definecolor{red1}{rgb}{0.95,0.9,0.9} 
\definecolor{red2}{rgb}{1,0.8,0.8} 
\definecolor{red3}{rgb}{1,0.6,0.6} 
\definecolor{red4}{rgb}{1,0,0} 

\definecolor{green1}{rgb}{0.9,0.95,0.9} 
\definecolor{green2}{rgb}{0.8,1,0.8} 
\definecolor{green3}{rgb}{0.6,1,0.6} 
\definecolor{green4}{rgb}{0,1,0} 

\definecolor{white}{rgb}{1,1,1} 
\definecolor{gray}{rgb}{0.75,0.75,0.75}
\newcommand{\ingray}[1]{\color{gray}#1\color{black}}

\begin{table}[htbp]
\centering\medskip
\begin{tabular}{rcccc}
& Induced & Did $\geq$ 5 & Fine & Average Hourly \\
& to work & labelings & Quality & Wage \\
\hline \hline
Meaningful & \cellcolor{green2} $\uparrow$ 4.6\%* & \cellcolor{green4} $\uparrow$ 8.5\%*** & \ingray{$\uparrow$ 0.7\%} & \ingray{$\downarrow$ 4.5\%} \\ 
Meaningful (US)  & \cellcolor{green2} $\uparrow$ 5.1\%* & \cellcolor{green3} $\uparrow$ 8.9\%** & \cellcolor{white} $\uparrow$ 3.9\% & $\downarrow$ 7.7\% \\ 
Meaningful (India)  & \ingray{$\downarrow$ 2.3\%} & \cellcolor{green2} $\uparrow$ 7.0\%* & \ingray{$\downarrow$ 3.1\%} & \ingray{$\uparrow$ 0.5\%} \\ \hline
Shredded & $\downarrow$ 4.0\% & \ingray{$\downarrow$ 2.8\%} & \cellcolor{red4} $\downarrow$ 7.2\%*** & \ingray{$\uparrow$ 5.6\%} \\ 
Shredded (US) & \ingray{$\downarrow$ 2.3\%} & \ingray{$\downarrow$ 5.0\%} & \cellcolor{red2} $\downarrow$ 6.1\%* & $\uparrow$ 9.5\% \\ 
Shredded (India)  & \cellcolor{white} $\downarrow$ 6.8\% & \ingray{$\downarrow$ 1.6\%} & \cellcolor{red3} $\downarrow$ 8.7\%** & \ingray{$\downarrow$ 1.4\%} \\ 
\hline\hline
\multicolumn{5}{p{.8\textwidth}}{\footnotesize * $p < .05$, \quad ** $p < .01$, \quad *** $p < .001$}
\end{tabular}
\caption{A heatmap illustration of our results. Rows 1 and 4 consider data from both America and India combined. Columns 1, 2, 3 show the results of regressions and column 4 shows the result of two-sample t-tests. Results reported are from regressions without demographic controls.}
\label{tab:effects_summary}
\end{table}

\subsection{Labor Participation Results: ``Induced to work''}

We investigate how treatment and country affects whether or not subjects chose to do our task. Unlike in a laboratory environment, our subjects were workers in a relatively anonymous labor market and were not paid a ``show-up fee.'' On MTurk, workers frequently start but do not finish tasks; attrition is therefore a practical concern for employers who hire from this market. In our experiment, on average, 25\% of subjects began, but did not follow-through by completing one full labeling.

Even in this difficult environment, we were able to increase participation among workers by roughly 4.6\% by framing the task as more meaningful (see columns 1 and 2 of table \ref{tab:quantity_results}). The effect is robust to including various controls for age, gender, and time of day effects. As a subject in the meaningful treatment told us, ``It's always nice to have [HITs] that take some thought and mean something to complete. Thank you for bringing them to MTurk.'' The shredded treatment discouraged workers and caused them to work 4.0\% less often but the effect was less significant ($p = 0.057$ without controls and $p = 0.082$ with controls). Thus, hypothesis 1 seems to be correct.

\begin{table}
\small
\begin{tabular}{l*{6}{c}}
\hline\hline
            &\multicolumn{1}{c}{Induced}&\multicolumn{1}{c}{Induced}&\multicolumn{1}{c}{Did $\geq$ 2}&\multicolumn{1}{c}{Did $\geq$ 2}&\multicolumn{1}{c}{Did $\geq$ 5}&\multicolumn{1}{c}{Did $\geq$ 5}\\
\hline
Meaningful     &       0.046*  &       0.046*  &       0.047*  &       0.050*  &       0.085***&       0.088***\\
            &     (0.020)   &     (0.020)   &     (0.022)   &     (0.022)   &     (0.024)   &     (0.024)   \\
Shredded    &       -0.040   &      -0.037   &      -0.012   &      -0.005   &      -0.028   &      -0.023  \\
            &     (0.021)   &     (0.021)   &     (0.022)   &     (0.022)   &     (0.024)   &     (0.024)   \\ \hline
India       &      -0.185***&      -0.183***&      -0.170***&      -0.156***&      -0.035   &      -0.003 \\
            &     (0.017)   &     (0.018)   &     (0.018)   &     (0.019)   &     (0.019)   &     (0.021)   \\
Male         &               &      0.006   &               &      -0.029   &               &      -0.081***\\
            &               &    (0.018)   &               &     (0.019)   &               &     (0.021)   \\
Constant      &       0.848***&       0.907***&       0.785***&       0.873***&       0.387***&       0.460***\\
\hline
Controls &&&&&& \\
~~Age     &               &        0.23   &               &        0.29   &               &        0.92   \\
~~Time of Day     &               &        0.16   &               &        0.06   &               &        0.46   \\
~~Day of Week     &               &        0.08   &               &        0.00**   &               &        0.55   \\
\hline
$R^2$          &        0.05   &        0.06   &        0.04   &        0.05   &        0.01   &        0.02   \\
$N$           &     2471   &     2471   &     2471   &     2471   &     2471   &     2471   \\
\hline\hline
\multicolumn{7}{p{.8\textwidth}}{\footnotesize * $p < .05$, \quad ** $p < .01$, \quad *** $p < .001$}
\end{tabular}
\caption{Robust linear regression results for the main treatment effects on quantity of images. Columns 1, 3 and 5 only include treatments and country. Columns 2, 4, and 6 control for gender, age categories, time of day, and day of week. Rows 6-8 show $p$-values for the partial $F$-test for sets of different types of control variables.}
\label{tab:quantity_results}
\end{table}

Irrespective of treatment, subjects from India completed an image 18.5\% less often ($p < 0.001$) than subjects from the US. We were interested in interactions between country and treatment, so we ran the separate induced-to-work regression results by country (unshown). We did not find significant effects within the individual countries because we were underpowered to detect this effect when the sample size was halved. We find no difference in the treatment effect for induced to work between India and the United States ($p=0.97$). This is inconsistent with hypothesis 4 where we predicted Indian subjects to respond more strongly to pecuniary incentives.

It is also possible that the effects for induced to work were weak because subjects could have still attributed meaning to the zero context and shredded conditions, a problem that will affect our results for quantiy and quality as well. This serves to bias our treatment effects downward suggesting that the true effect of meaning would be larger. For instance, one zero-context subject told us, ``I assumed the `objects' were cells so I guess that was kind of interesting.'' Another subject in the zero-context treatment advised us, ``you could put MTurkers to good use doing similar work with images, e.g. in dosimetry or pathology ... and it would free up medical professionals to do the heftier work.''

\subsection{Quantity Results: Number of images labeled}

Table \ref{tab:summary_stats} shows that the number of images increased with meaning. However, this result is conditional on being induced to work and is therefore contaminated with selection bias. We follow \citet{Angrist2001} and handle selection by creating a dummy variable for ``did two or more labelings'' and a dummy for ``did five or more labelings'' and use them as responses (other cutoffs produced similar results).

We find mixed results regarding whether the the level of meaningfulness affects the quantity of output. Being assigned to the meaningful treatment group \textit{did} have a positive effect, but assignment to the shredded treatment did not result in a corresponding decrease in output.

Analyzing the outcome ``two or more labelings,'' column 3 of table \ref{tab:quantity_results} shows that the meaningful treatment induced 4.7\% more subjects to label two or more images ($p < 0.05$). The shredded treatment had no effect. Analyzing the outcome ``five or more labelings'' (column 5), which we denote as ``high-output workers,''\footnote{Labeling five or more images corresponds to the top tercile of quantity among people who were induced to work.} the meaningful treatment was highly significant and induced 8.5\% more workers ($p < 0.001$ with and without controls), an increase of nearly 23 percent, and the shredded treatment again has no effect.

Hypothesis 2 (quantity increases with meaningfulness) seems to be correct only when comparing the meaningful treatment to the zero-context treatment. An ambiguous effect of the shredded treatment on quantity is also reported by \citet{Ariely_search_for_meaning_2008}. 

We didn't find differential effects between the United States and India. In an unshown regression, we found that Americans were 9.5\% more likely to label five or more images ($p < 0.01$) and Indians were 8.4\% more likely to label five or more ($p < 0.05$). These two effects were not found to be different ($p = 0.84$) which is inconsistent with hypothesis 4 that Indians are more motivated by pecuniary incentives than Americans.

Interestingly, we also observed a number of ``target-earners'' who stopped upon reaching exactly one dollar in earnings. A mass of 16 participants stopped at one dollar, while one participant stopped at \$1.02 and not one stopped at \$0.98, an effect also observed by \citet{horton_targetearners_2010}. The worker who labored longest spent 2 hours and 35 minutes and labeled 77 images.

\subsection{Quality Results: Accuracy of labeling}

Quality was measured by the fraction of cells labeled at a distance of five pixels (``coarse quality'') and two pixels (``fine quality'') from their true centers. In presenting our results (see table \ref{tab:quality_results}), we analyze the treatment effects using our fine quality measure. The coarse quality regression results were similar, but the fine quality had a much more dispersed distribution.\footnote{The inter-quartile range of coarse quality overall was [93.3\%,  97.2\%] whereas the IQR of fine quality was overall [54.7\%, 80.0\%].} 

\begin{table}[htp]
\footnotesize
\begin{tabular}{l*{6}{c}}
\hline\hline
& \multicolumn{6}{c}{Fine Quality} \\ 
& \multicolumn{2}{c}{Both Countries} & \multicolumn{2}{c}{United States} & \multicolumn{2}{c}{India} \\ \hline
Meaningful     &       0.007   &       0.014   &       0.039   &       0.039*  &      -0.031   &      -0.013   \\
                    &    (0.017)   &     (0.014)   &     (0.023)   &     (0.019)   &     (0.025)   &     (0.021)    \\
Shredded      &      -0.072***&      -0.074***&      -0.061*  &      -0.066** &      -0.087** &      -0.073**  \\
            &    (0.021)   &     (0.017)   &     (0.027)   &     (0.023)   &     (0.031)   &     (0.023)   \\ \hline
India       &       -0.053***&      -0.057***&               &               &               &               \\
            &     (0.015)   &     (0.013)   &               &               &               &               \\
Male         &               &       0.053***&               &       0.014   &               &       0.100***\\
            &               &     (0.013)   &               &     (0.017)   &               &     (0.021)    \\
Labelings 6---10    &               &      -0.018** &               &      -0.024** &               &      -0.016*   \\
            &               &     (0.006)   &               &     (0.008)   &               &     (0.008)  \\
Labelings $\geq$ 11   &               &      -0.140***&               &      -0.116***&               &      -0.148*** \\
            &               &     (0.017)   &               &     (0.029)   &               &     (0.020)  \\
Constant &      0.666***&       0.645***&       0.651***&       0.625***&       0.634***&       0.588***\\
\hline
Controls &&&&&&\\
~~Image     &               &        0.00***   &               &        0.00***   &               &        0.00*** \\
~~Age     &               &        0.10   &               &        0.01**   &               &        0.25   \\
~~Time of Day      &               &        0.33   &               &        0.29   &               &        0.78   \\
~~Day of Week     &               &       0.12   &               &        0.46   &               &        0.26   \\
\hline
$R^2$          &        0.04   &        0.15   &        0.04   &        0.12   &        0.02   &        0.20   \\
$N$           &    12724   &    12724   &     6777   &     6777   &     5947   &     5947   \\
\hline\hline
\multicolumn{7}{p{.8\textwidth}}{\footnotesize * $p < .05$, \quad ** $p < .01$, \quad *** $p < .001$}
\end{tabular}
\caption{Robust linear regression clustered by subject for country and treatment on fine quality as measured by the number of cells found two pixels from their exact centers.  Columns 1, 3 and 5 include only treatments and country. Columns 2, 4, and 6 control for number of images, the particular image (of the ten images), gender, age categories, time of day, and day of week.}
\label{tab:quality_results}
\end{table}

Our main result is that fine quality was 7.2\% lower in the shredded treatment, but there wasn't a large corresponding increase in the meaningful treatment.\footnote{One caveat with our quality results is that we only observe quality for people who were induced to work and selected into our experiment (we have ``attrition bias''). Attrition was  4\% higher in the shredded treatment and we presume that the people who opted out of labeling images would have labeled them with far worse quality had they remained in the experiment.} This makes sense; if the workers knew their labelings weren't going to be checked, there is no incentive to mark points carefully. This result was not different across countries (regression unshown). The meaningful treatment has a marginally significant effect only in the United States, where fine quality increased by 3.9\%  ($p=0.092$ without controls and $p = 0.044$ with controls), but there was no effect in India. Thus, hypothesis 3 (quality increases with meaningfulness) seems to be correct \textit{only} when comparing the shredded to the zero context treatment which is surprising.

Although Indian workers were less accurate than United States workers and had 5.3\% lower quality ($p < 0.001$ and robust to controls), United States and Indian workers did not respond differentially to the shredded treatment ($p=0.53$). This again is inconsistent with hypothesis 4.

Experience matters. Once subjects had between 6 and 10 labelings under their belt, they were 1.8\% less accurate ($p < 0.01$), and if they had done more than 10 labelings, they were 14\% less accurate ($p<0.001$). This result may reflect negative selection --- subjects who labeled a very high number of images were probably working too fast or not carefully enough.\footnote{Anecdotally, subjects from the shredded condition who submitted comments regarding the task were less likely to have expressed concerns about their accuracy. One subject from the meaningful group remarked that ``[his] mouse was too sensitive to click accurately, even all the way zoomed in,'' but we found no such apologies or comments from people in the shredded group.} Finally, we found that some of the ten images were substantial harder to label accurately than others (a partial F-test for equality of fixed effects results in $p < 0.001$).

\subsection{Post Manipulation Check Results}\label{subsec:pmc}

In order to understand how our treatments affected the perceived meaningfulness of the task, we gave a post manipulation check to all subjects who completed at least one image and did not abandon the task before payment. This data should be interpreted cautiously given that subjects who completed the tasks and our survey are \textit{not} representative of all subjects in our experiment.\footnote{Ideally, we would have collected this information immediately after introducing the treatment condition. However, doing so would have compromised the credibility of our natural field experiment.}

We found that those in the meaningful treatment rated significantly higher in the post manipulation check in both the United States and India. Using a five-point Likert scale, we asked workers to rate the perceived level of meaningfulness, purpose, enjoyment, accomplishment, and recognition. In the meaningful treatment, subjective ratings were higher in all categories but the self-rated level of meaningfulness and purpose were the highest. The level of meaningfulness was 1.3 points higher in the US and 0.6 points higher in the India; the level of perceived porposefulness was 1.2 points higher in America and 0.5 points higher in India. In the United States, the level of accomplishment only increased by 0.8 and the level of enjoyment and recognition increased by 0.3 and 0.5 respectively with a marginal increase in India. As a US participant told us, ``I felt it was a privilege to work on something so important and I would like to thank you for the opportunity.'' 

We conclude that the meaningful frames accomplished their goal. Remarkably, those in the shredded treatment in either country did not report significantly lower ratings on any of the items in the post manipulation check. Thus, the shredded treatment may not have had the desired effect.

\section{Conclusion}\label{sec-conclusion}

Our experiment is the first that uses a natural field experiment in a real labor market to examine how a task's meaningfulness influences labor supply.

Overall, we found that the greater the amount of meaning, the more likely a subject is to participate, the more output they produce, the higher quality output they produce, and the less compensation they require for their time. We also observe an interesting effect: high meaning increases \textit{quantity} of output (with an insignificant increase in quality) and low meaning decreases \textit{quality} of output (with no change in quantity). It is possible that the level of perceived meaning affects how workers substitute their efforts between task quantity and task quality. The effect sizes were found to be the same in the US and India.

Our finding has important implications for those who employ labor in any short-term capacity besides crowdsourcing, such as temp-work or piecework. As the world begins to outsource more of its work to anonymous pools of labor, it is vital to understand the dynamics of this labor market and the degree to which non-pecuniary incentives matter. This study demonstrates that they do matter, and they matter to a significant degree.

This study also serves as an example of what MTurk offers economists: an excellent platform for high internal validity natural field experiments while evading the external validity problems that may occur in laboratory environments.

\bibliographystyle{elsarticle-harv}
\bibliography{refs}

\pagebreak
\appendix

\section{Detailed Experimental Design}\label{sec:exp_design_appendix}

\begin{figure}[htp]
\centering
\includegraphics[width=4in]{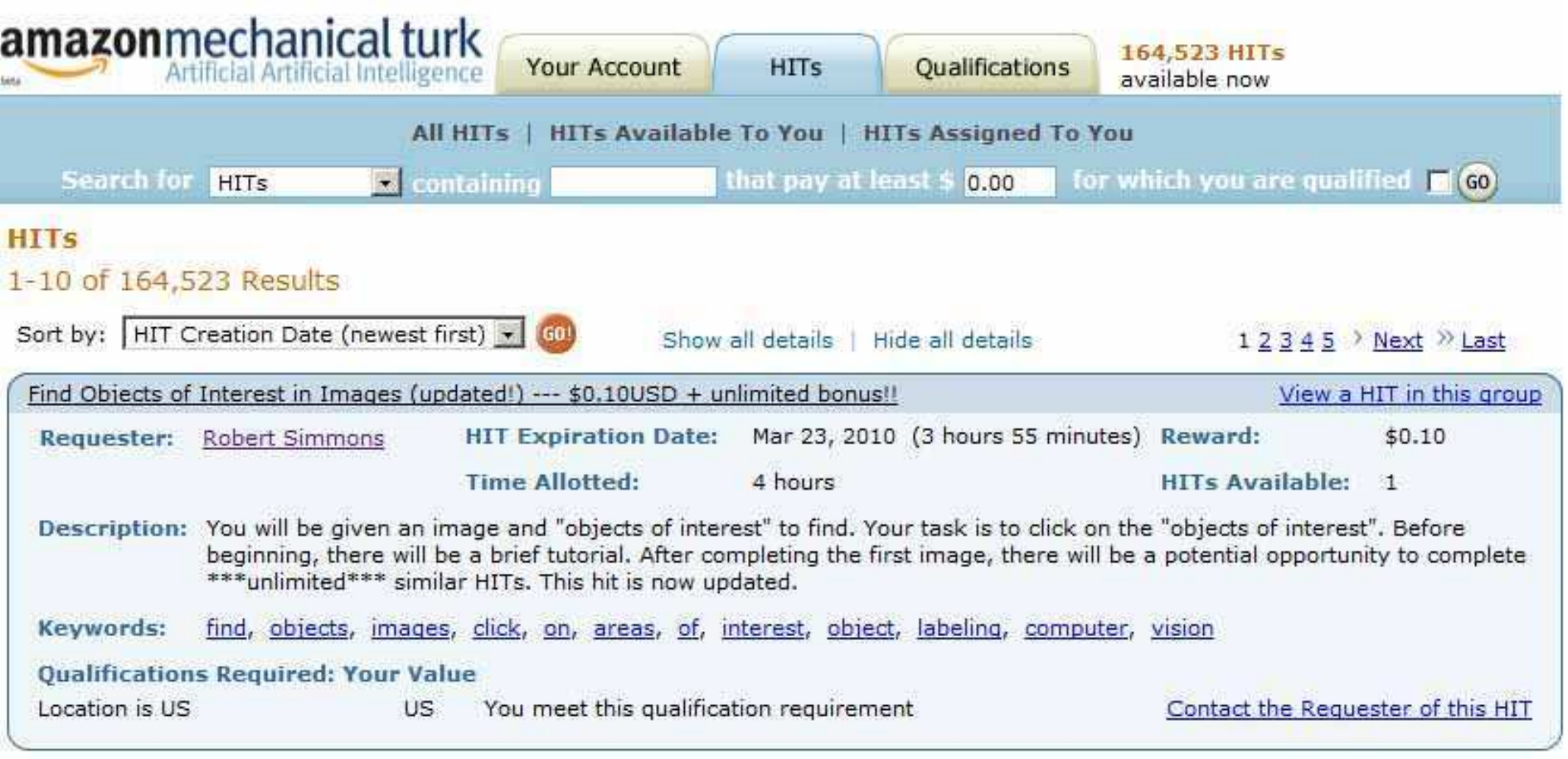}
\caption{The HIT as initially encountered on MTurk. Note: we used an alias in order to appear as a non-corporate and non-institutional employer.}
\label{fig:app_just_hit_and_amazon}
\end{figure}

This section details exact screens shown to users in the experimental groups. The worker begins by encountering the HIT on the MTurk platform (see Figure \ref{fig:app_just_hit_and_amazon}).

\begin{figure}[htp]
\centering
\includegraphics[width=4in]{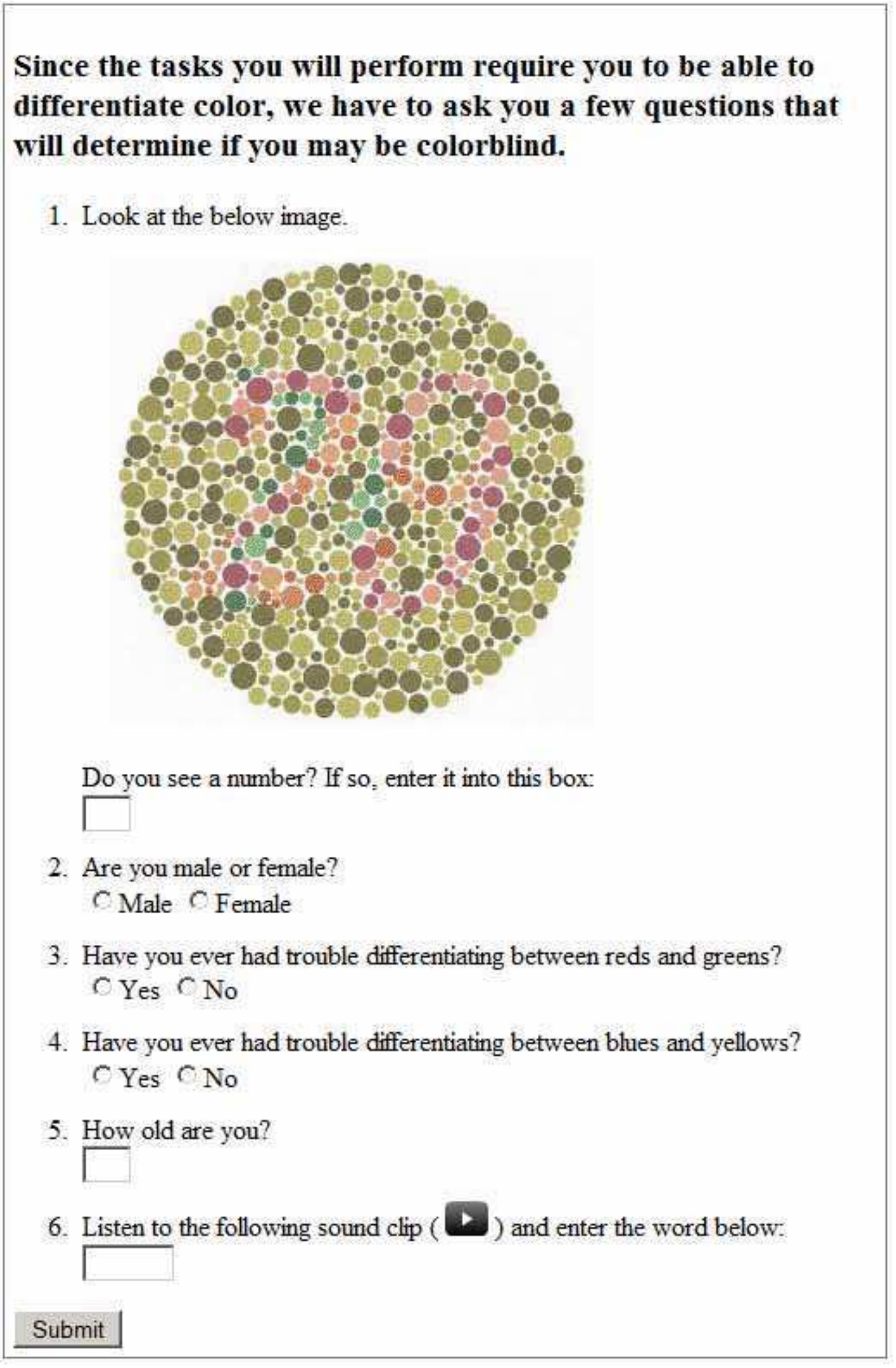}
\caption{The colorblindness test}
\label{fig:app_colorblindness}
\end{figure}

The worker can then click on the HIT and they see the ``preview screen'' which describes the HIT (not shown) with text. In retrospect, a flashy image enticing the worker into the HIT would most likely have increased throughput. If the worker chooses to accept, they are immediately directed to a multi-purpose page which hosts a colorblindness test, demographic survey, and an audio test for functioning speakers (see Figure \ref{fig:app_colorblindness}). Although many tasks require workers to answer questions before working, we avoided asking too many survey-like questions to avoid appearing as an experiment. 

At this point, the worker is randomized into one of the three treatments and transitioned to the ``qualification test.''  The page displays an instructional video varying by treatment which they cannot fast-forward. Screenshots of the video are shown in Figures \ref{fig:app_video_1}, \ref{fig:app_video_2}, and \ref{fig:app_video_3}.\footnote{We thank Rob Cohen who did an excellent job narrating both scripts.}

We include the verbatim script for the videos below. Text that differs between treatments is typeset in square brackets separated by a slash. The text before the slash in red belongs to the meaningful treatment and the text following the slash in blue belongs to both the zero-context and shredded treatments. 

\begin{quote}
\scriptsize

Thanks for participating in this task. [\color{red} Your job will be to help identify tumor cells in images and we appreciate your help. / \color{blue} In this task, you'll look at images and find objects of interest.] \color{black} 

In this video tutorial, we'll explain [\color{red} three / \color{blue} two] \color{black} things: 

[\color{red} First, why you're labeling the images, which is to help researchers identify tumorous cancer cells. Next, we'll show you how to identify those tumor cells. / \color{blue} First, we'll show you how to identify objects of interest in images.] [\color{red} Finally, / \color{blue} Then,] \color{black} we'll explain how after labeling your first image you'll have a chance to label some more.

\begin{figure}[htp]
\centering
\subfigure[Zero-context / Shredded treatments]{
\includegraphics[width=2.0in]{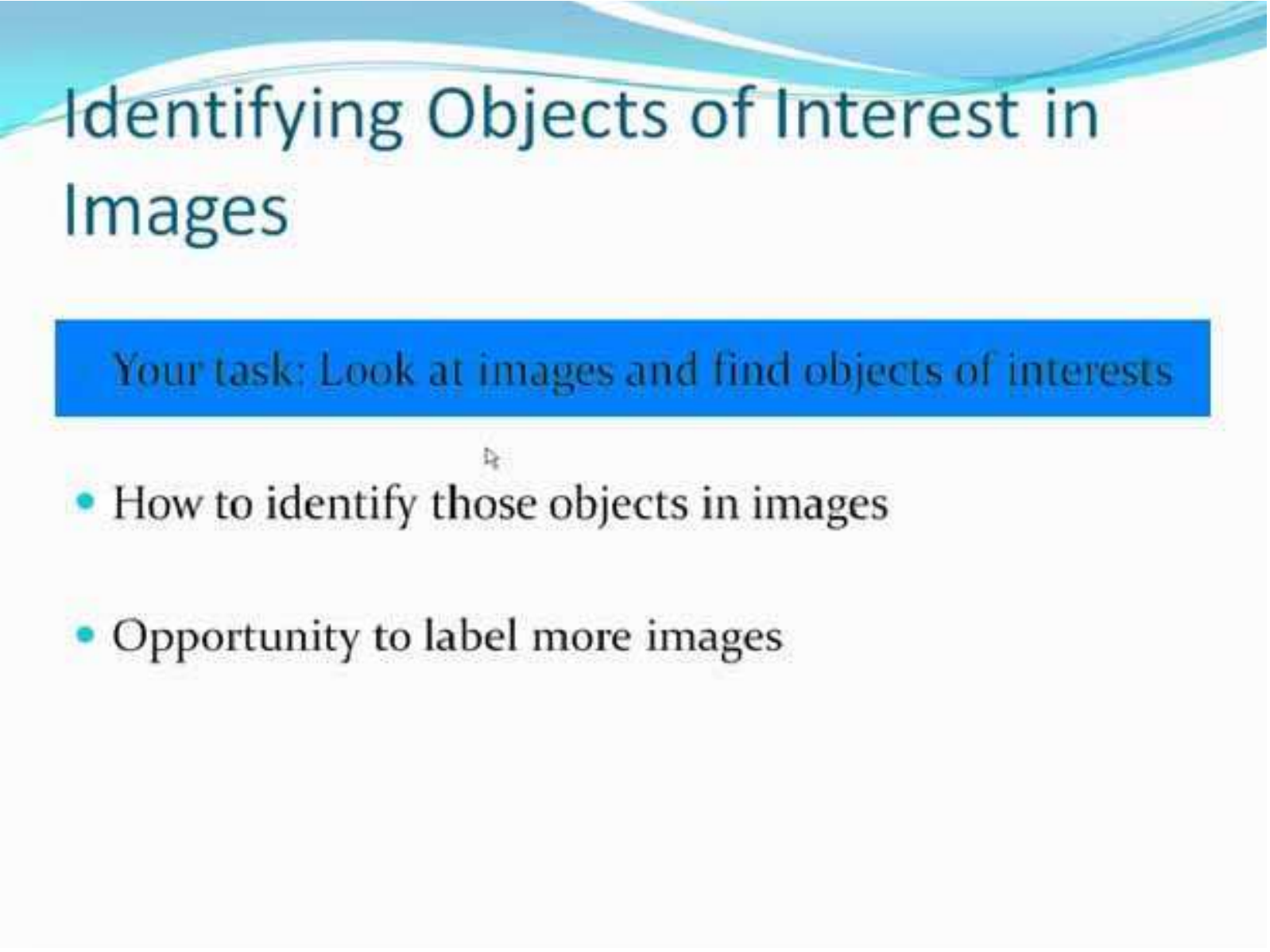}
}
\subfigure[Meaningful treatment]{
\includegraphics[width=2.0in]{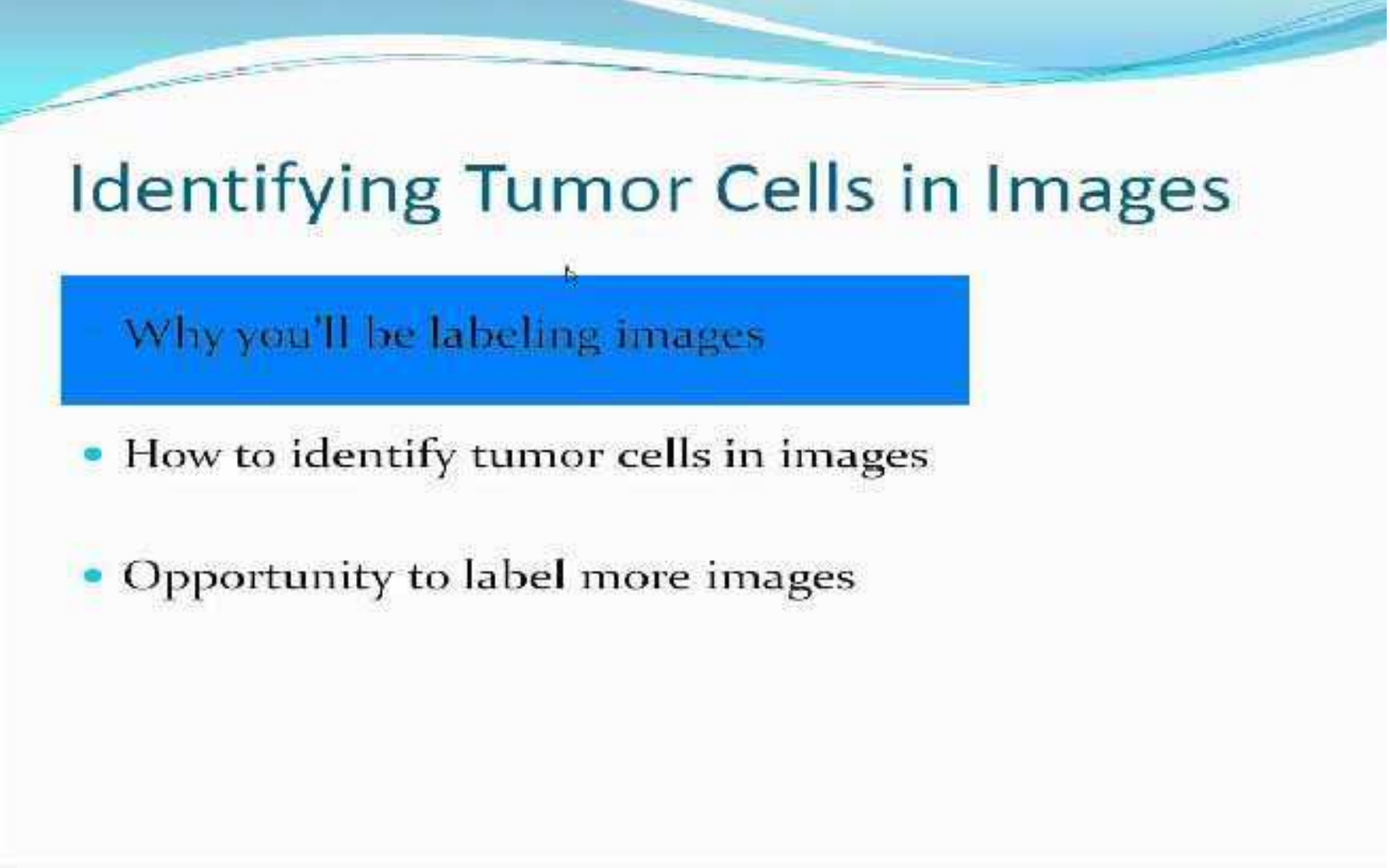}
}
\caption{Opening screen of training video.}
\label{fig:app_video_1}
\end{figure}

\begin{figure}[htp]
\centering
\begin{minipage}[b]{.45\textwidth}
\includegraphics[width=2.0in]{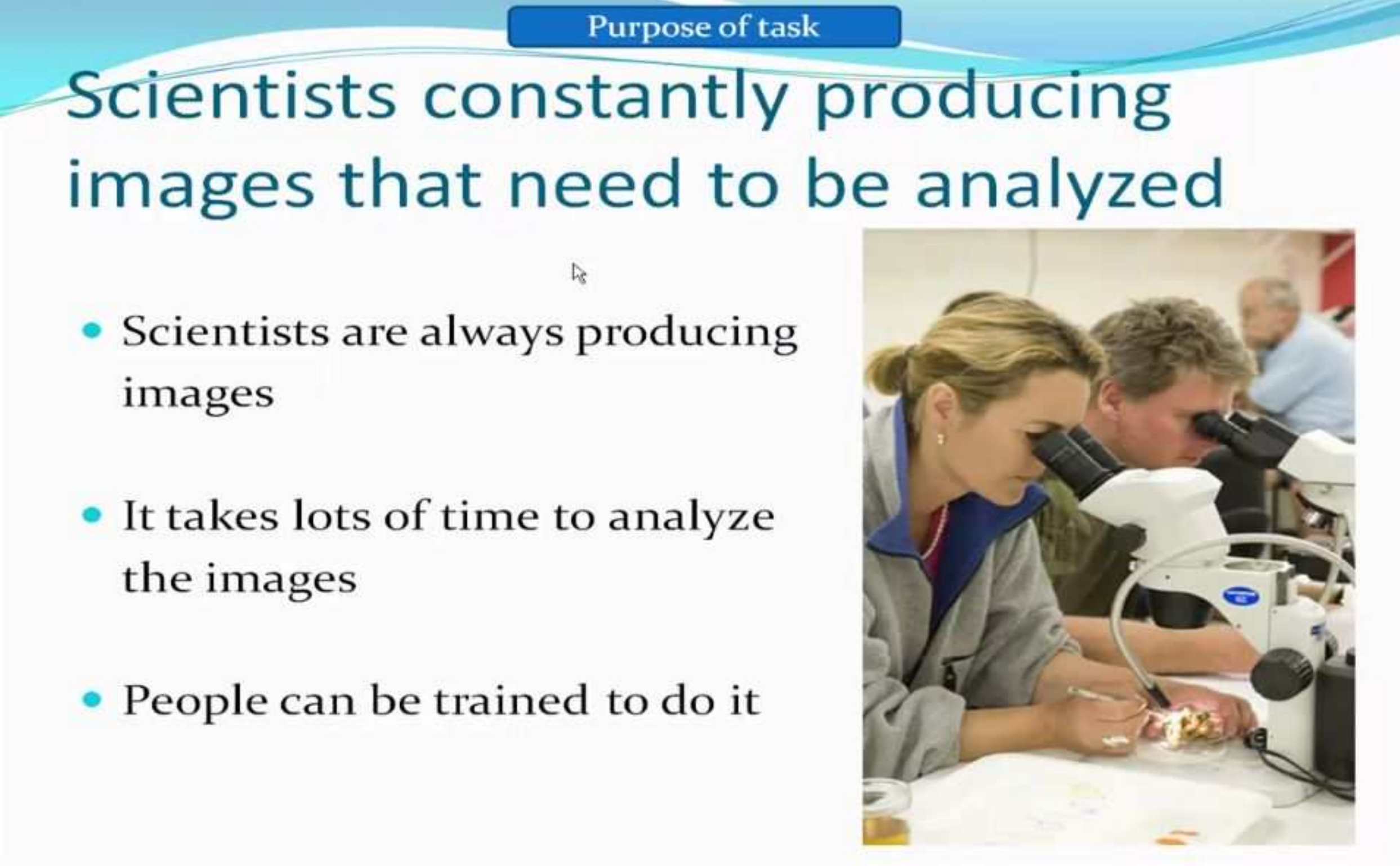}
\end{minipage}
\begin{minipage}[b]{.45\textwidth}
\includegraphics[width=2.0in]{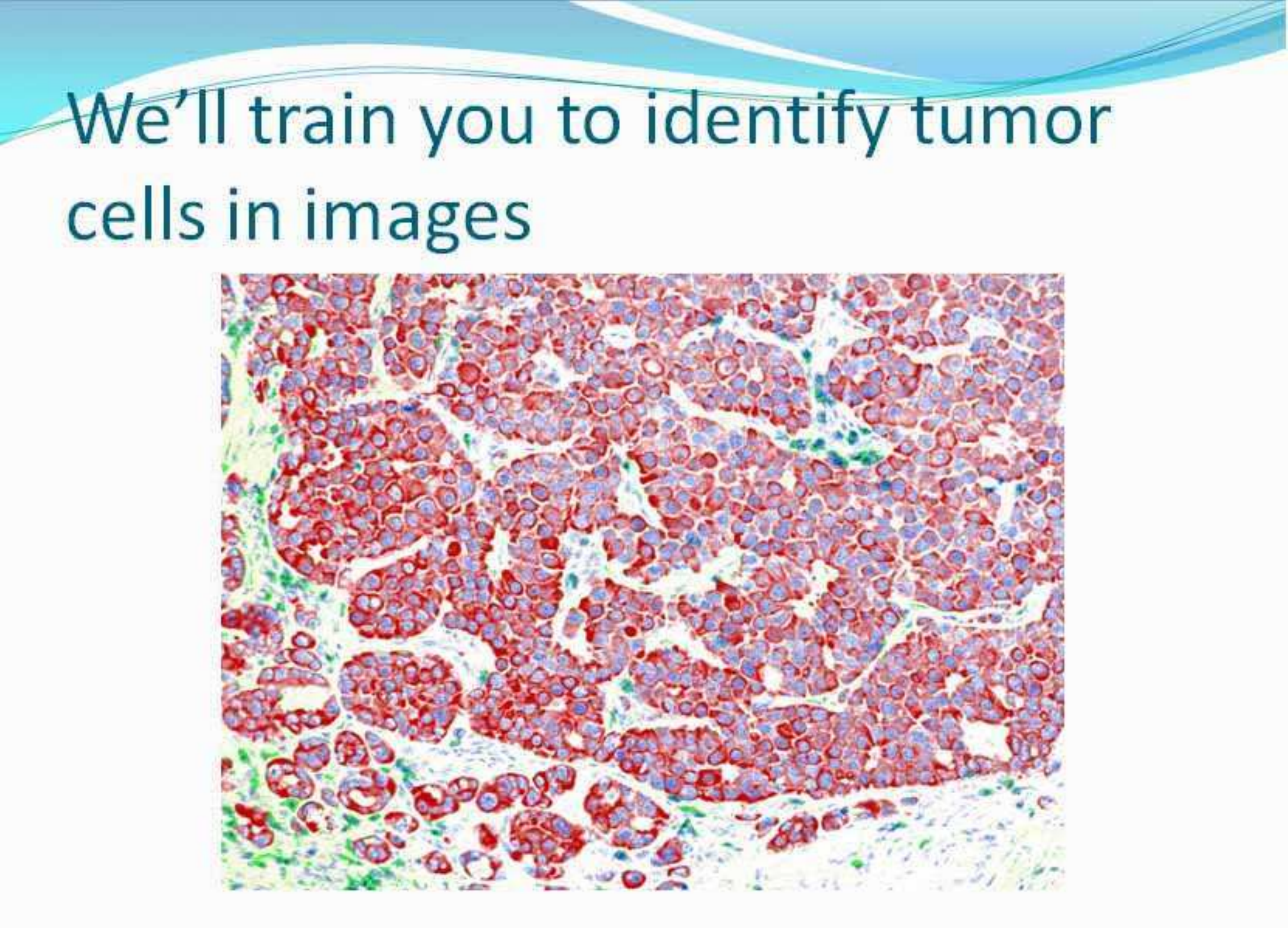}
\end{minipage}
\caption{Examples of meaningful cues which are not present in the Zero-context and Shredded treatment instructional video.}
\label{fig:app_video_2}
\end{figure}

Now we're ready to learn how to identify [\color{red} tumor cells / \color{blue} objects of interest] \color{black} in images. Some example pictures of the [\color{red} tumor cells / \color{blue} objects of interest] \color{black} you'll be identifying can be found at the bottom left. Each [\color{red} tumor cell / \color{blue} object of interest] \color{black} is blue and circular and surrounded by a red border. 

\begin{figure}[htp]
\centering
\subfigure[Zero-context / Shredded treatments]{
\includegraphics[width=2.0in]{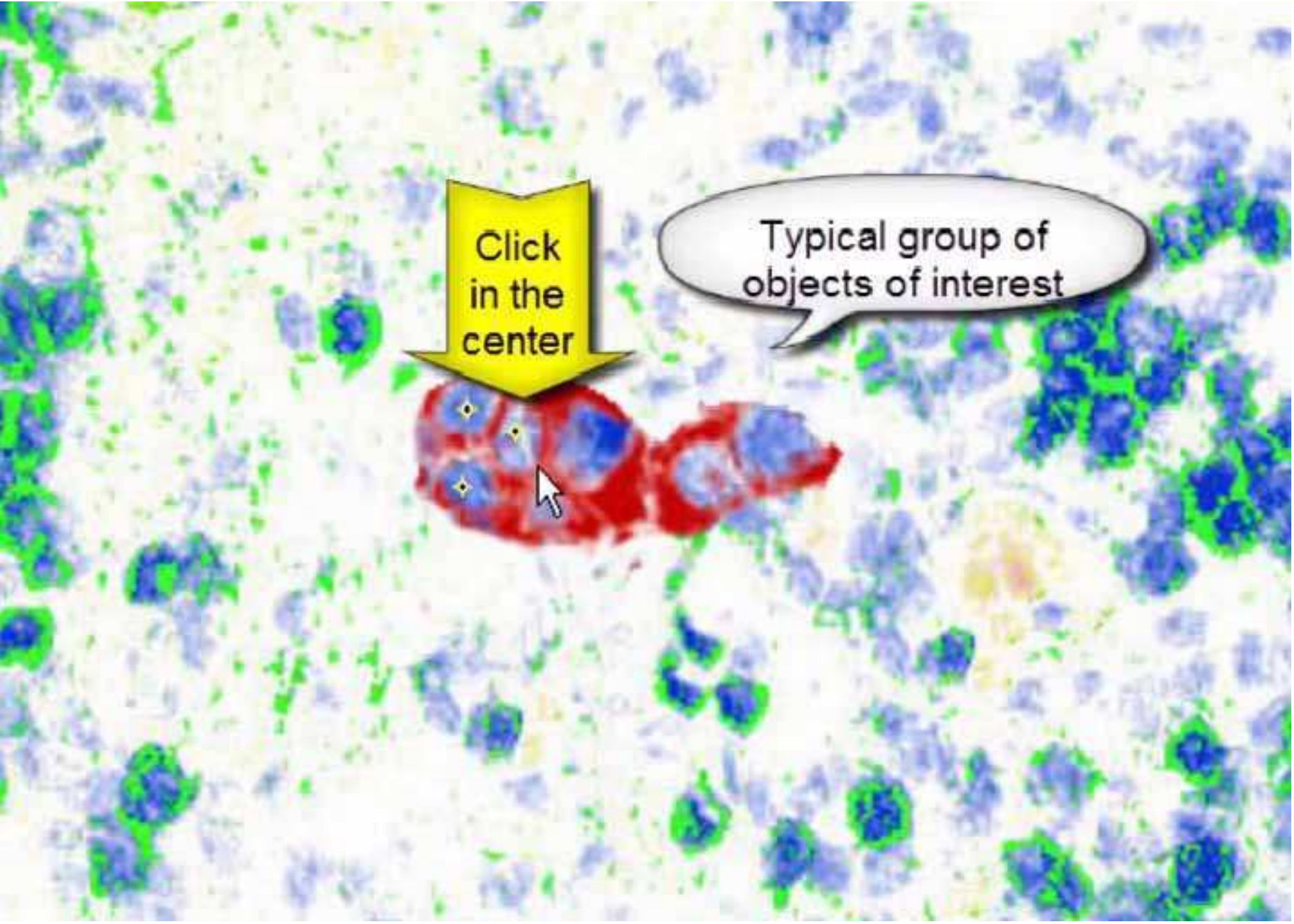}
}
\subfigure[Meaningful treatment]{
\includegraphics[width=2.0in]{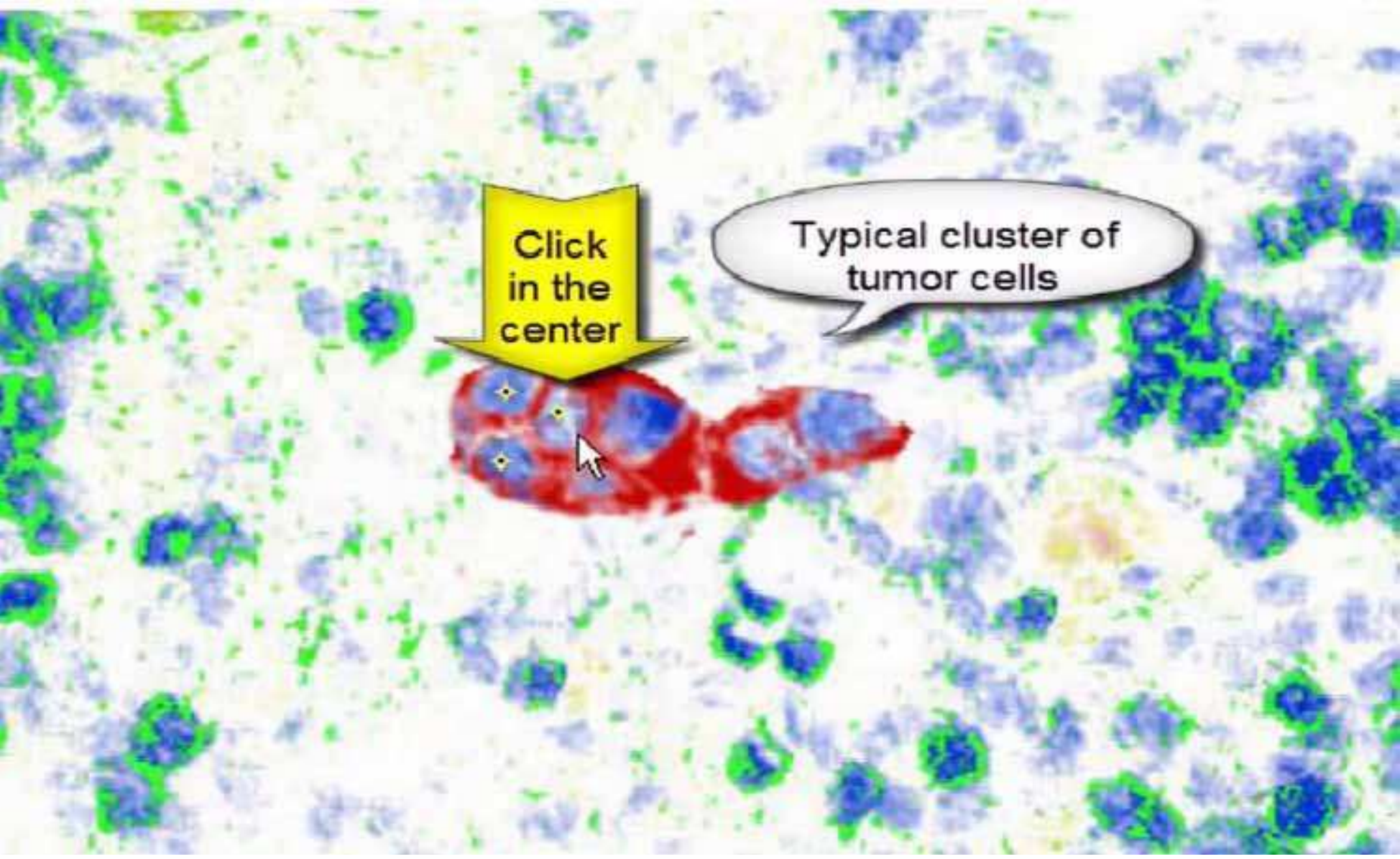}
}
\caption{Describing the training process.}
\label{fig:app_video_3}
\end{figure}

When you begin each image, the magnification will be set to the lowest resolution. This gives you an overview of all points on the image, but you'll need to zoom in and out in order to make the most precise clicks in the center of the [\color{red} tumor cells / \color{blue} objects of interest]\color{black}.

Let's scroll through the image and find some [\color{red} tumor cells / \color{blue} objects of interest] \color{black} to identify.

Here's a large cluster of [\color{red} tumor cells / \color{blue} objects of interest]\color{black}. To identify them, it is very important to click as closely to the center as possible on each [\color{red} cell / \color{blue} object] \color{black}. If I make a mistake and don't click in the center, I can undo the point by right-clicking. 

Notice that this [\color{red} cell / \color{blue} point] \color{black} isn't entirely surrounded by red, [\color{red} probably because the cell broke off]\color{black}. Even though it's not entirely surrounded by red, we still want to identify it as a [\color{red} tumor cell / \color{blue} object of interest]\color{black}.

In order to ensure that you've located all [\color{red} tumor cells / \color{blue} objects of interest]\color{black}, you should use the thumbnail view in the top right. You can also use the magnification buttons to zoom out. 

It looks like we missed a cluster of [\color{red} tumor cells / \color{blue} objects of interest] \color{black} at the bottom. Let's go identify those points.

Remember once again, that if you click on something that is not a [\color{red} tumor cell / \color{blue} object of interest]\color{black}, you can unclick by right-clicking. 

Using the scroll bars, we'll navigate to the other points ... and here's some more to the left ... Now that we think we've identified all points, let's zoom out to be sure and scroll around.

Before submitting, we should be sure of three things: (1) That we've identified all [\color{red} tumor cells / \color{blue} objects of interest]\color{black}~(2) That we've clicked in the center of each one (3) That we haven't clicked on anything that's not a [\color{red} tumor cell / \color{blue} object of interest]\color{black}.

Once we've done that, we're ready to submit.

Finally, after you complete your first image, you'll have an opportunity to label additional images as part of this HIT.

The first images you label will pay more to compensate for training.

After that, as part of this HIT you'll have the chance to identify as many additional images as you like as long as you aren't taking more than 15 minutes per image.

Although you can label unlimited images in this HIT, you won't be able to accept more HITs. This is to give a variety of turkers an opportunity to identify the images.

[\color{red} Thank you for your time and effort. Advances in the field of cancer and treatment prevention rely on the selfless contributions of countless individuals such as yourself.\color{black}]
\end{quote}

\begin{figure}[htp]
\centering
\includegraphics[width=4in]{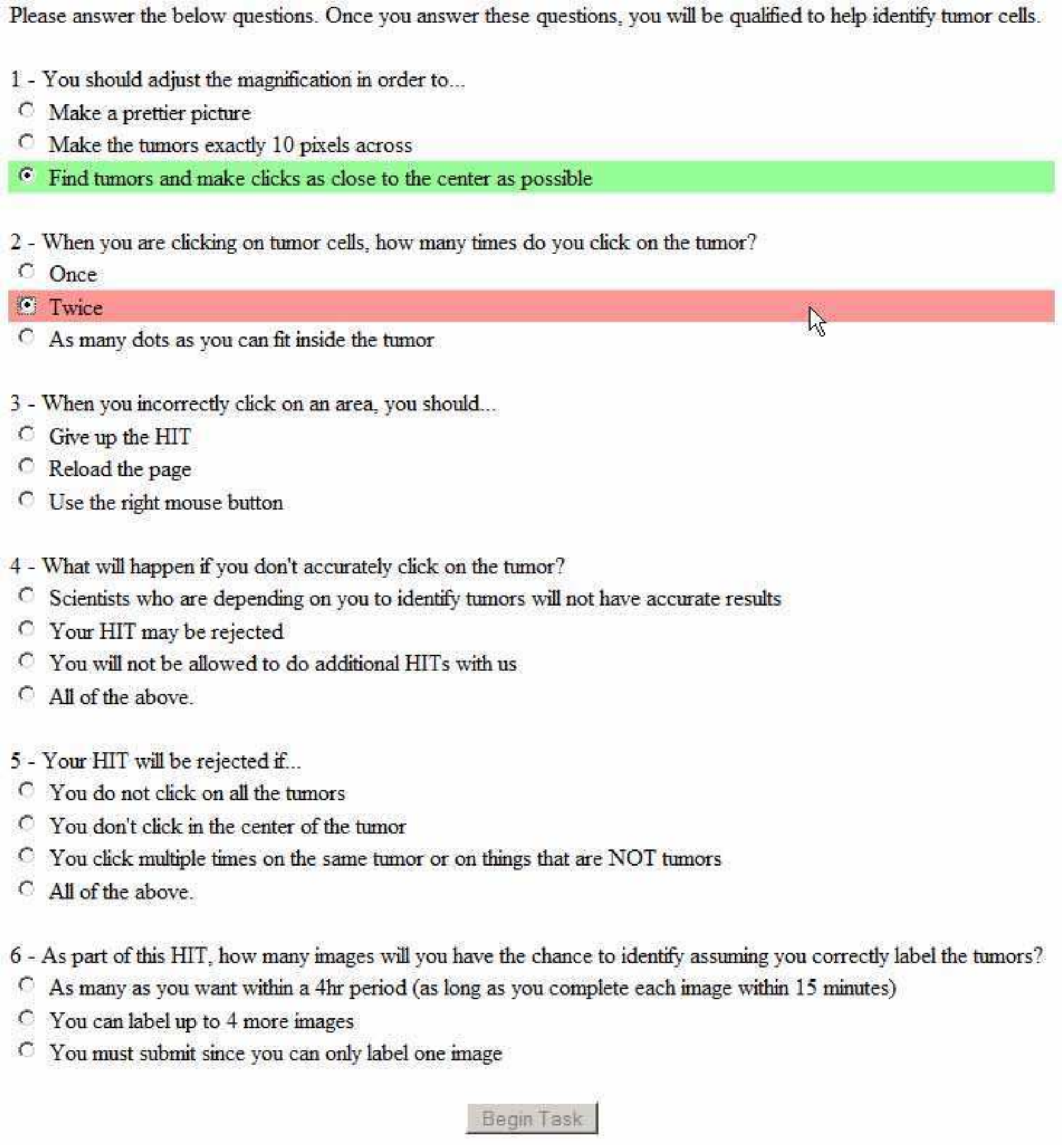}
\caption{The quiz after watching the training video for the meaningful treatment. In the zero-context and shredded treatments, all mention of ``tumor cells'' are replaced by ``objects of interest.'' The shredded treatment has an additional question asking them to acknowledge that they are working on a test system and their work will be discarded. Green indicates a correct response; red indicates an incorrect response.}
\label{fig:app_quiz}
\end{figure}

Then, workers must take a quiz (see Figure \ref{fig:app_quiz}). During the quiz, they can watch the video freely (which was rarely done).

\begin{figure}[htp]
\centering
\subfigure[Meaningful treatment]{
\includegraphics[width=2in]{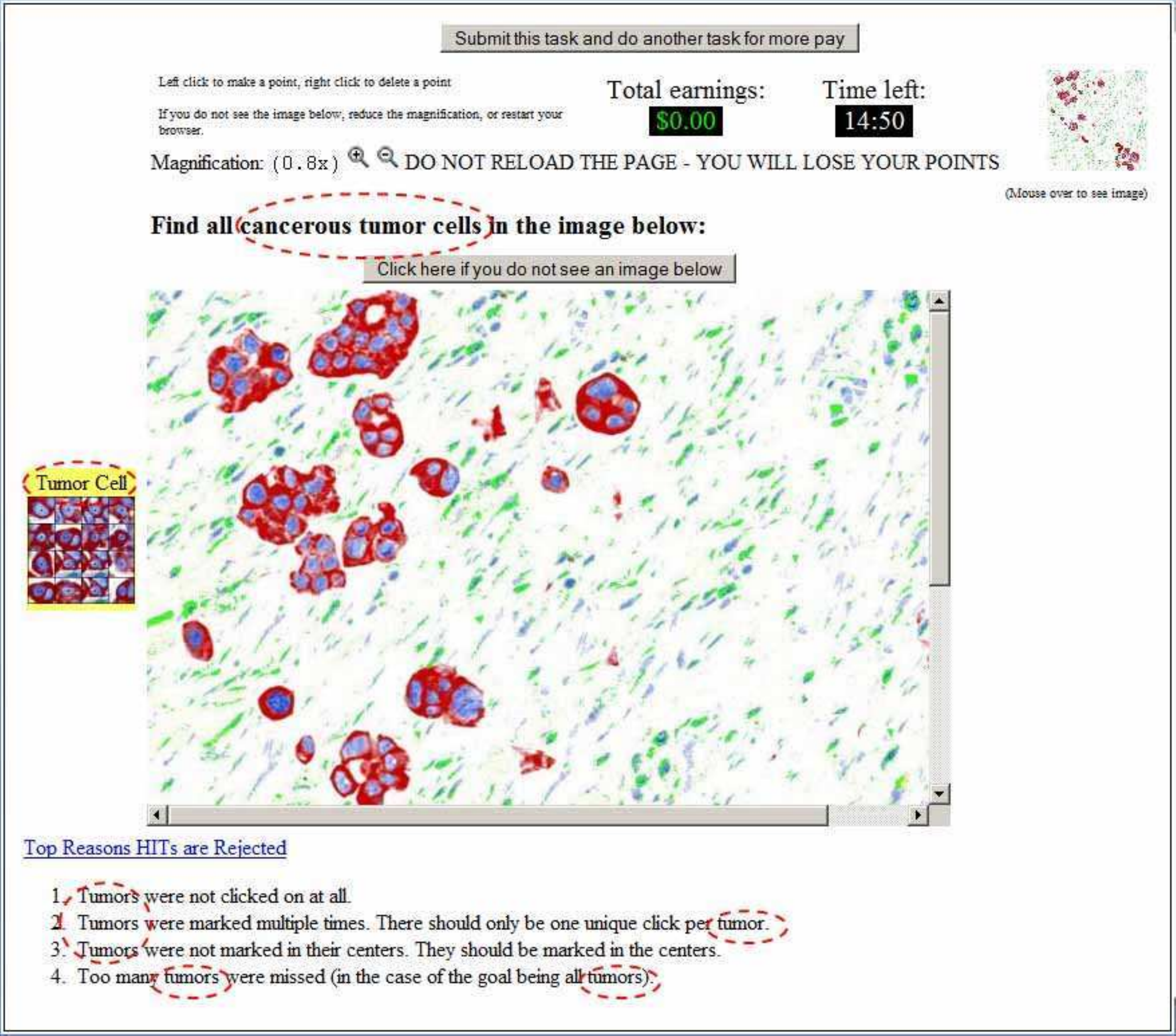}
}
\subfigure[Zero-context / Shredded treatments]{
\includegraphics[width=2in]{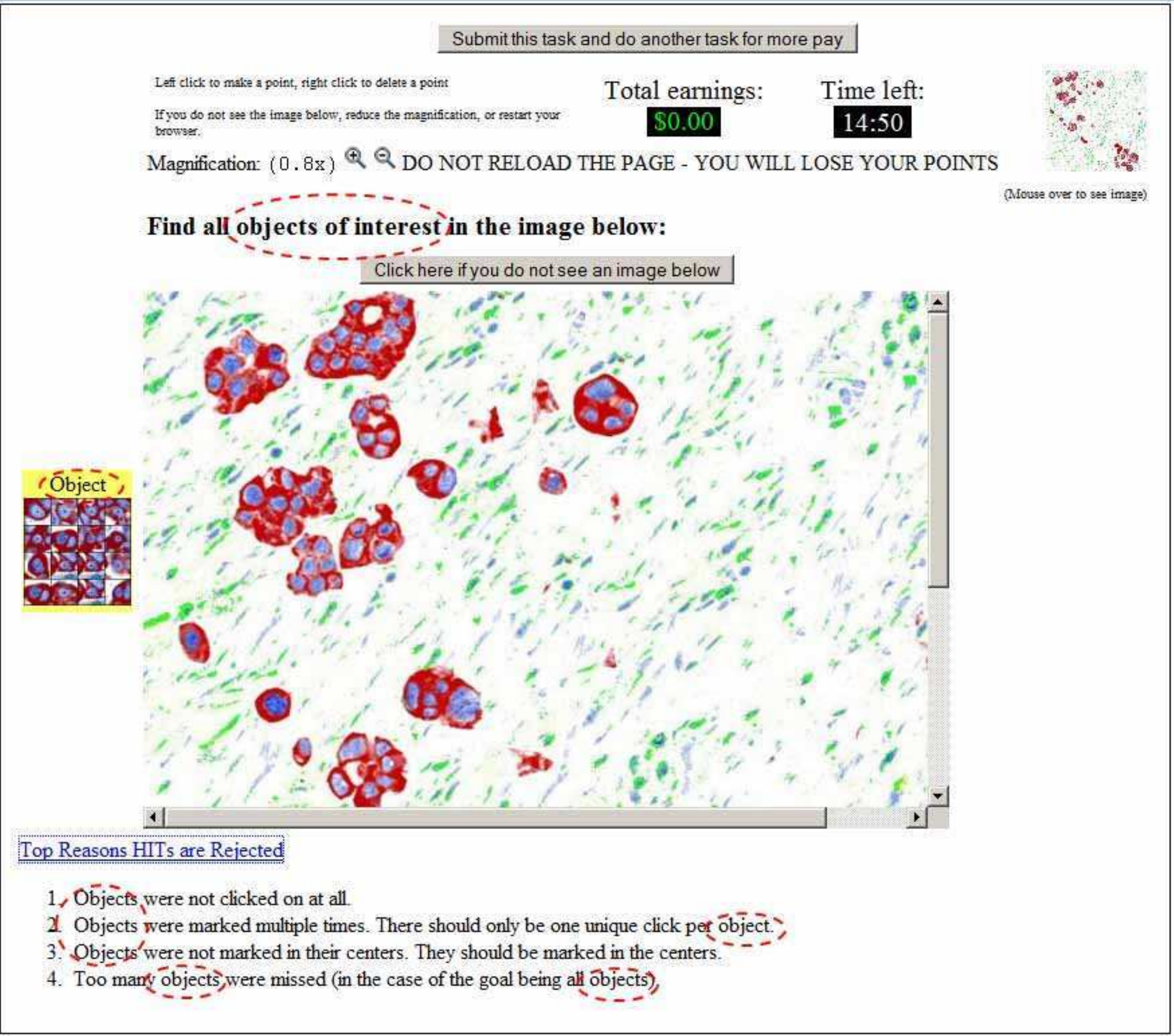}
}
\caption{The training interface as seen by workers. The meaningful interface reminds the subjects in 8 places that they are identifying tumor cells. The zero-context interface only says ``objects of interest'' and the shredded condition in addition has a message in red indicating that their points will not be saved (unshown). The circles around each point were \textit{not} visible to participants. We display them to illustrate the size of a 10-pixel radius.}
\label{fig:app_interface}
\end{figure}

Upon passing, they began labeling their first image (see Figure \ref{fig:app_interface}). The training interface includes the master training window where workers can create and delete points and scroll across the entire image. To the left, there is a small image displaying example tumor cells. Above the master window, they have zoom in / out buttons. And on the top right there is a thumbnail view of the overall image.

Participants were given 15 minutes to mark an image. Above the training window, we displayed a countdown timer that indicated the amount of time left. The participant's total earnings was also prominently displayed atop. On the very top, we provided a submit button that allowed the worker to submit results at any time.

Each image had the same 90 cells from various-sized clusters. The cell clusters were selected for their unambiguous examples of cells, thereby eliminating the difficulty of training the difficult-to-identify tumor cells. In each image, the same clusters were arranged and rotated haphazardly, then pasted on one of five different believable backgrounds using Adobe Photoshop. Those clusters were then further rotated to create a set of ten images. This setup guarantees that the difficulty was relatively the same image-image. Images were displayed in random order for each worker, repeating after each set of ten (repetition was not an issue since it was rare for a participant to label more than ten).

\begin{figure}[htp]
\centering
\subfigure[Zero-context / Shredded treatments]{
\includegraphics[width=2.0in]{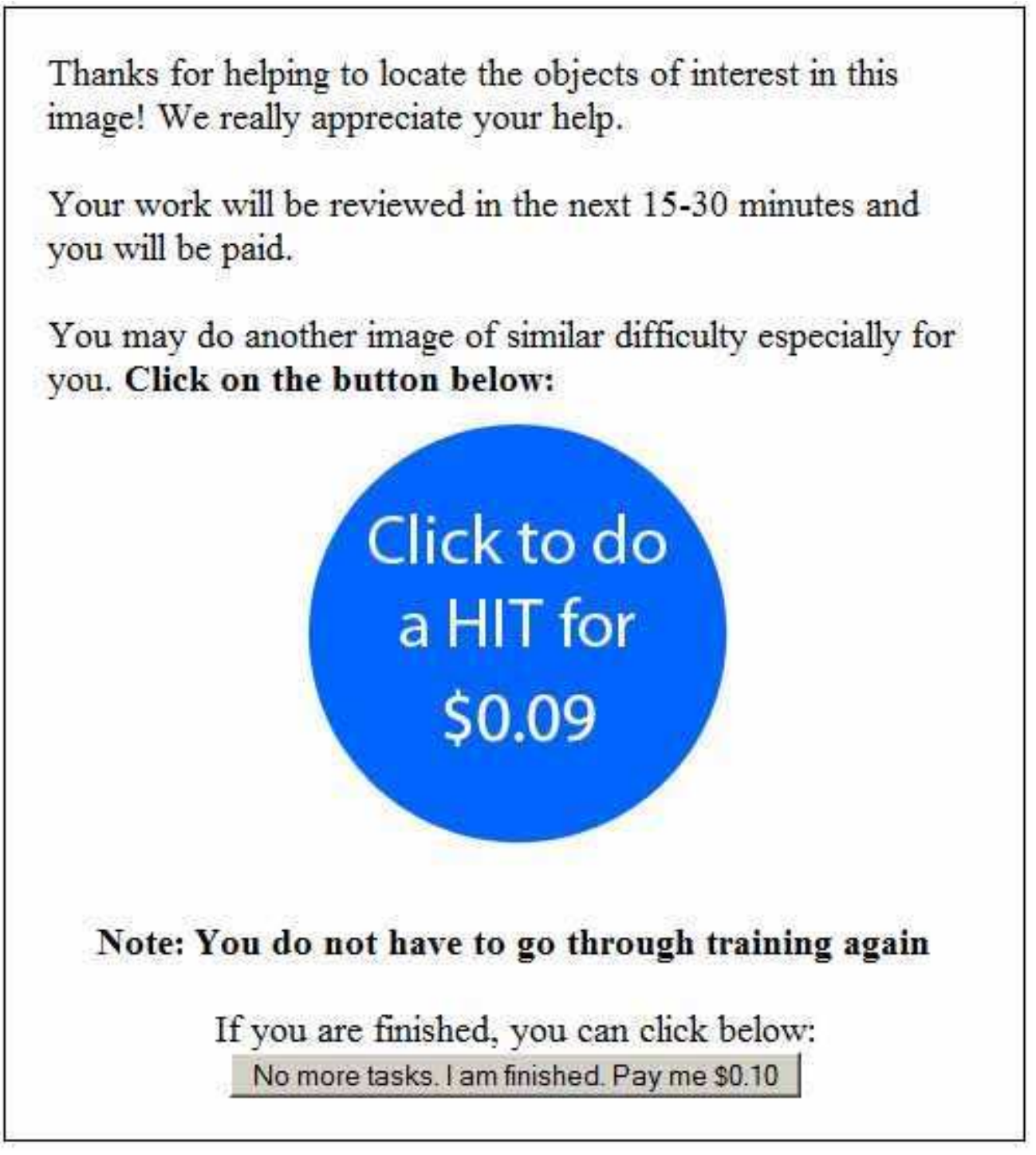}
}
\subfigure[Meaningful treatment]{
\includegraphics[width=2.0in]{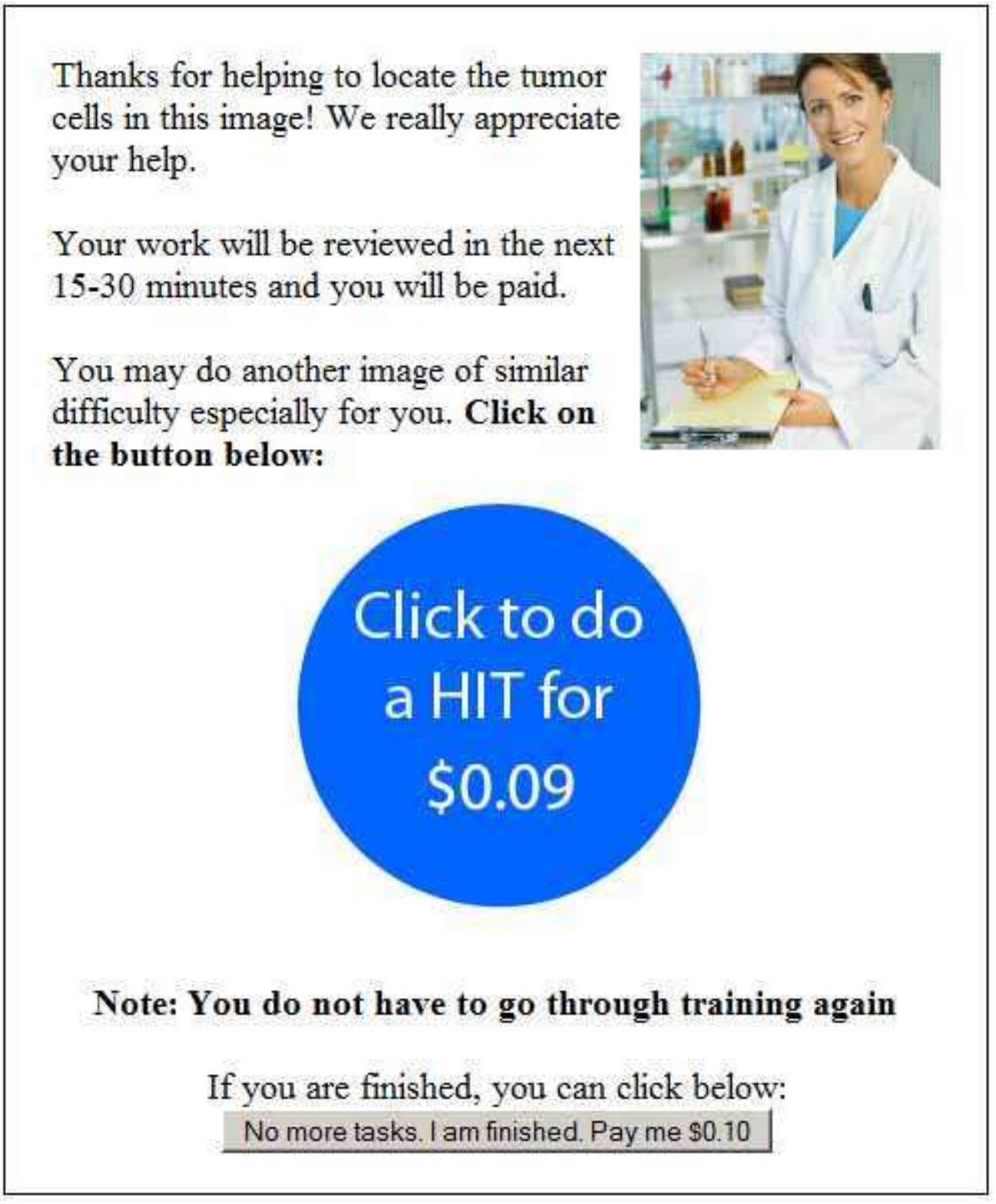}
}
\caption{The landing page after a labeling task is completed. At this point, workers are asked if they'd like to label another image or quit and be paid what they've earned so far.}
\label{fig:app_do_another}
\end{figure}

\begin{figure}[htp]
\centering
\includegraphics[width=5in]{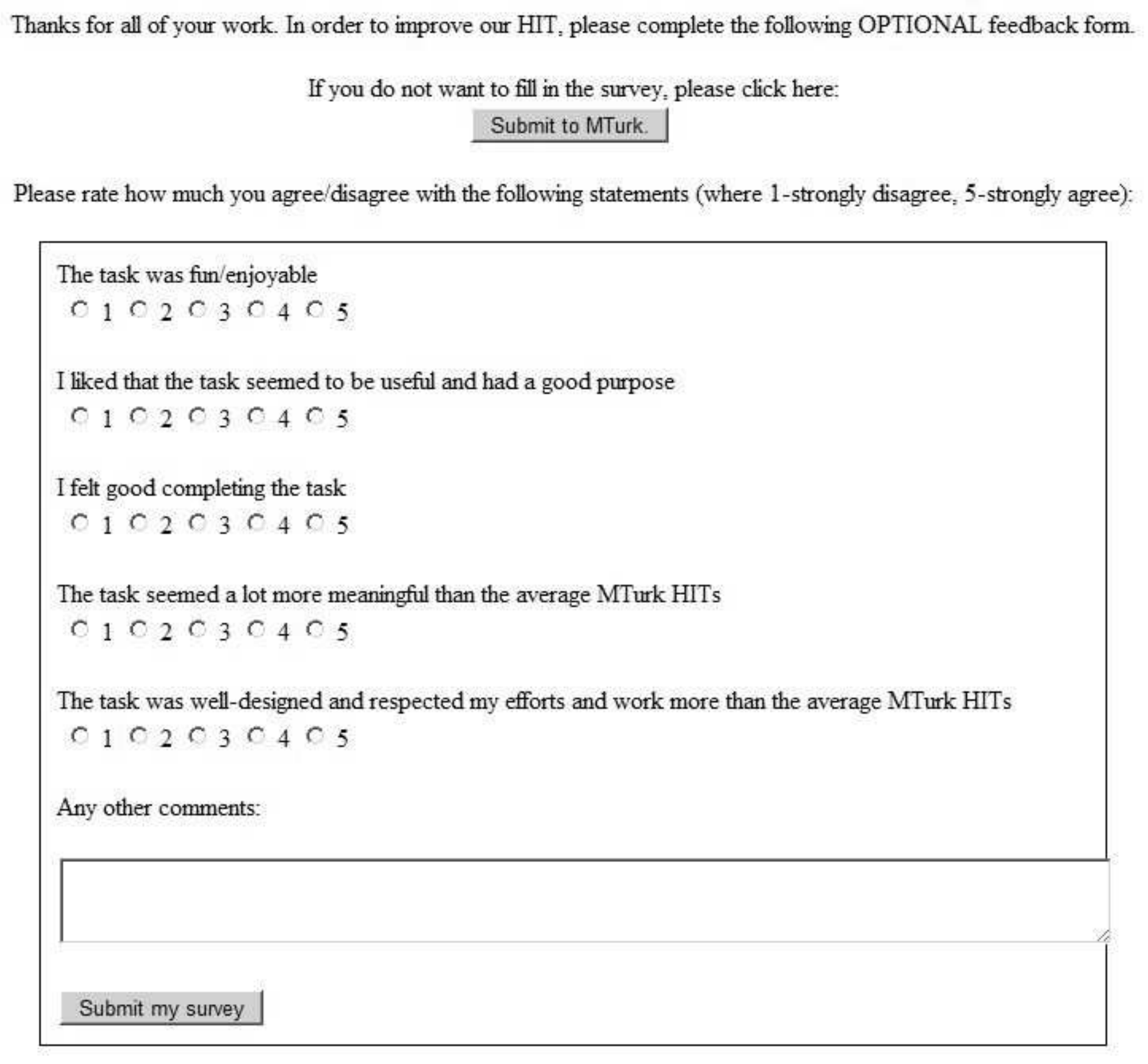}
\caption{The survey a subject fills out upon completion of the task.}
\label{fig:post_survey}
\end{figure}

After the worker is finished labeling, the worker presses submit and they are led to an intermediate page which asks if they would like to label another image and the new wage is prominently displayed (see Figure \ref{fig:app_do_another}). In the meaningful treatment, we add one last cue of meaning --- a stock photo of a researcher to emphasize the purpose of the task. In the shredded treatment, we append the text ``NONE of your points will be saved because we are testing our system, but you will still be paid.'' If the worker wishes to continue, they are led to another labeling task; otherwise, they are directed to the post manipulation check survey shown in figure \ref{fig:post_survey}.

The program ensures that the worker is being honest. We require them to find more than 20\% of the cells (the workers were unaware that we were able to monitor their accuracy). If they are found cheating on three images, they are deemed \textit{fraudulent} and not allowed to train more images. Since payment is automatic, this is to protect us from a worker depleting our research account. In practice, this happened rarely and was not correlated with treatment.

\section{A Technical Guide to Running Field Experiments on Mechanical Turk}\label{sec:technical_discussion}

\subsection*{Institutional Review Board (IRB) Requirements}

This study requires the use of deception in order to observe social preferences in a natural environment and thus is not exempted under category 2's survey procedures. The issue is you cannot give the subjects an initial consent form indicating that they are part of an experiment. 

Upon waiving the requirement of consent, the IRB will most likely require you to issue a debrief statement to your subjects stating that they were part of an experiment, a blurb about the purpose of the experiment, and contact information to your institution's IRB. In order for the experiment to work properly, you can only issue the debriefing \textit{after} data collection is completed. Otherwise, MTurk subjects can communicate to each other that this is a study and this may be a problem for internal validity.

\subsection*{Engineering Required}

The implementation of an experiment of similar scale to the one we describe in the paper requires no more than two weeks of full-time work for an experienced software engineer.

It is critical that the engineer be fluent in ``front-end'' design. The front-end is what your subjects will use throughout the experiment and it can be highly dynamic, responding to the individual participant's actions. MTurk tasks are rendered in HTML and CSS. Javascript controls the dynamism and AJAX provides smooth client-server communication. Although not used in this study, the front-end can become even more fancy by implementing Adobe Flash or Microsoft Silverlight applets.

An MTurk experiment also requires a back-end web stack consisting of an http server, a database, and a server-side platform. The back-end's function is to render webpages and store the experiment's data. We recommend Ruby on Rails 3.1 because of its rapid development speed, inexpensive and instant deployment to rented space in ``the Cloud,'' and because it is free and open-source. 

Since the server must communicate with MTurk, it is convenient for the engineer to also have experience using the Amazon web services application programming interface (MTurk's API) as well as CRON jobs on Linux.

We recommend setting up two cron jobs for effective experimentation:

\begin{enumerate}
\item A CRON that creates HITs every 15 minutes. In order for subjects to see your task, it must remain fresh on Amazon's main page. Creating tasks every 15 min with short expiration times (an hour or so) will allow for maximum throughput.
\item Another CRON that automatically pays workers. We recommend doing this every hour or so. It's important that this be done automatically including the bonuses and rejections otherwise it will be cumbersome for the experimenter to check through each and every assignment.
\end{enumerate}

Apart from technical skills, the engineer should also have knowledge of the principles of experimental design.

\end{document}